\documentclass{mn2e}
\usepackage{graphicx}
\usepackage{amsbsy}
\usepackage{amsmath}
\usepackage{epsf}
\usepackage{paralist}
\usepackage{upgreek}
\usepackage{tipa}
\usepackage[inline]{enumitem}
\usepackage{arydshln}
\usepackage{amssymb}
\usepackage{pifont}
\usepackage{epstopdf}

\newcommand{\norm}[1]{\lvert #1 \rvert}

\newcommand{\beq}{\begin{equation}}
\newcommand{\eeq}{\end{equation}}

\DeclareMathAlphabet{\mathsfsl}{OT1}{cmss}{bx}{sl}
\SetMathAlphabet{\mathsfsl}{bold}{OT1}{cmss}{bx}{sl}

\newcommand{\xmark}{\ding{55}}%

\newcommand{\RN}[1]{%
  \textup{\uppercase\expandafter{\romannumeral#1}}%
}

\def\mean#1{\left< #1 \right>}
\usepackage{color}        

\begin{document}

\title[non-ideal MHD effects in cores]
{Non-ideal MHD simulations of subcritical prestellar cores with non-equilibrium chemistry}

\author[Tritsis et al.]
  {A.~Tritsis$^{1, 2}$\thanks{E-mail: atritsis@uwo.ca}, C.~Federrath$^{1}$, K.~Willacy$^{3}$, K.~Tassis$^{4, 5}$ \\
    $^1$Research School of Astronomy and Astrophysics, Australian National University, Canberra, ACT 2611, Australia \\
    $^2$Department of Physics and Astronomy, University of Western Ontario, London, ON N6A 3K7, Canada \\
    $^3$MS 169-507, Caltech/Jet Propulsion Laboratory, 4800 Oak Grove Drive, Pasadena, CA 91109, USA \\
    $^4$Department of Physics and ITCP, University of Crete, Voutes, 70013 Heraklion, Greece \\
    $^5$Institute of Astrophysics, Foundation for Research and Technology-Hellas, Voutes, 70013 Heraklion, Greece}
\maketitle 

\begin{abstract}

Non-ideal magnetohydrodynamic (MHD) effects are thought to be gravity's closest ally in overcoming the support of magnetic fields and in forming stars. Here, we modify the publicly available version of the adaptive mesh refinement code \textsc{FLASH} (Fryxell et al. 2000; Dubey et al. 2008) to include a detailed treatment of non-ideal MHD and study such effects in collapsing prestellar cores. We implement two very extended non-equilibrium chemical networks, the largest of which is comprised of $\sim$ 300 species and includes a detailed description of deuterium chemistry. The ambipolar-diffusion, Ohmic and Hall resistivities are then self-consistently calculated from the abundances of charged species. We present a series of 2-dimensional axisymmetric simulations where we vary the chemical model, cosmic-ray ionization rate, and grain distribution. We benchmark our implementation against ideal MHD simulations and previously-published results. We show that, at high densities ($n_{\rm{H_2}}>~10^6~\rm{cm^{-3}}$), the ion that carries most of the perpendicular and parallel conductivities is not $\rm{H_3^+}$ as was previously thought, but is instead $\rm{D_3^+}$.

\end{abstract}


\section{Introduction}\label{intro}

Ever since their discovery (Hall 1949; Hiltner 1949), interstellar magnetic fields were recognized to play a pivotal role in the star-formation process (Mestel \& Spitzer 1956). For instance, the Galactic magnetic field is thought to be responsible for triggering large-scale instabilities which lead to the formation of molecular clouds (Mouschovias et al. 2009; S{\'a}nchez-Salcedo \& Santill{\'a}n 2011; Rodrigues et al. 2016; K{\"o}rtgen et al. 2018). Additionally, magnetic braking remains to date the only physical mechanism proposed to quantitatively resolve the angular momentum problem during the formation and contraction of molecular clouds (Mouschovias \& Paleologou 1979; Mouschovias \& Paleologou 1980; Basu \& Mouschovias 1994). Moreover, fragmentation initiated by ambipolar diffusion is also proposed as an explanation of the initial core mass function and consequently the initial mass function (IMF; Kunz \& Mouschovias 2009). Finally, numerical simulations (Gerrard et al. 2019) also revealed that ordered magnetic fields are a prerequisite for the formation of jets from protostellar discs (Frank et al. 2014).

A growing number of recent observations highlighted anew the importance of the magnetic field. Dense ``filamentary" structures are systematically observed to be oriented perpendicular to the magnetic field which is also found to be aligned with low-density structures (Planck Collaboration et al. 2016). At prestellar and protostellar cores scales, polarimetric observations with the Stratospheric Observatory for Infrared Astronomy (SOFIA) revealed ``hourglass" configurations of the magnetic field hinting towards a magnetically regulated star-formation scenario (Chuss et al. 2019; Redaelli et al. 2019). At these scales, the magnetic field is also expected to scale with the density of the cloud as $\rm{B}$$\propto$$\rm{\rho^{\upkappa}}$ (Mestel 1966; Mouschovias 1976b). The value of the exponent $\rm{\upkappa}$ depends on the geometry of the cloud and the direction of the contraction with respect to the mean magnetic field and could therefore reveal important information regarding the dynamical significance of the latter. However, the exact value of $\rm{\upkappa}$ still remains a topic of active debate and research (Crutcher et al. 2010; Tritsis et al. 2015; Jiang et al. 2020).

Be that as it may, the general consensus is that the magnetic field reduces the star-formation efficiency by (at least) a factor of a few by supporting molecular clouds and cores against their own self-gravity (Padoan \& Nordlund 2011; Federrath \& Klessen 2012; Federrath 2015). Specifically, the role of the magnetic field with respect to the self-gravity of the cloud is quantified through the mass-to-flux ratio with the central, critical value for collapse given by:
\begin{equation}\label{mtfc}
\bigg(\frac{dM}{d\rm{\Phi}}\bigg)_{c, cr} = \frac{3}{2}\bigg(\frac{1}{63\ G}\bigg)^{1/2}
\end{equation}
(Mouschovias \& Spitzer 1976; Mouschovias 1976a). If the mass-to-flux ratio exceeds this critical value then a cloud can overcome the support of the magnetic pressure and tension and collapse, even if ideal magnetohydrodynamic (MHD) conditions apply. Even in such a scenario however, ``flux freezing" has to break down at some stage in the evolution of the core. Otherwise the star that would form from such a core would have $\ge$4 orders of magnitude stronger magnetic field than that measured in some of the most magnetized stars ever observed (Wade et al. 2012). On the other hand, if a core is ``magnetically subcritical", $\upmu$ = $(\rm{dM}/\rm{d\Phi})\big/(\rm{dM}/\rm{d\Phi})_{c, cr}$~$<$~1, then it can never collapse under ``flux freezing" since the magnetic flux is conserved and the mass-to-flux ratio can never exceed the critical value. Such fragments have been found in Zeeman-splitting observations of the hyperfine transitions of the $\rm{CN}$ molecule (Crutcher 1999; Falgarone et al. 2008). On the other hand, under non-ideal MHD conditions, ambipolar diffusion can redistribute the mass in the central magnetic-flux tubes and these fragments can collapse (Mouschovias \& Ciolek 1999).

Regardless of whether the initial mass-to-flux ratio of molecular clouds is greater or smaller than the critical value, non-ideal MHD effects cannot be ignored since the flux-freezing approximation (i.e. magnetic flux is conserved) is limited. Even in highly ionized plasmas, such as in solar prominences where the ionization fraction is $\chi_i$$\ge$$10^{-1}$ (Patsourakos \& Vial 2002), drift velocities of the order of $\sim$1~km~$\rm{s^{-1}}$ between species with different ionization states have been observed (Khomenko et al. 2016). In comparison, in the densest regions of molecular-cloud fragments or cores the ionization fraction is $\chi_i$=$10^{-6}$--$10^{-9}$ (Caselli et al. 1998; Williams et al. 1998; Bergin et al. 1999; Caselli et al. 2002; Hezareh et al. 2008; Goicoechea et al. 2009). It therefore becomes clear that in these dense cores non-ideal MHD effects are unavoidable.

To date, a number of non-ideal MHD simulations of collapsing prestellar cores and clouds have been performed (e.g. Nakano 1979; Paleologou \& Mouschovias 1983; Lizano \& Shu 1989; Mouschovias \& Morton 1992; Fiedler \& Mouschovias 1993; Ciolek \& Mouschovias 1994; Basu \& Mouschovias 1995; Desch \& Mouschovias 2001; Tassis \& Mouschovias 2005; Duffin \& Pudritz 2008; Basu et al. 2009; Kunz \& Mouschovias 2010; Tassis et al. 2012; Marchand et al. 2016; Christie et al. 2017; Zhao et al. 2020). With the exception of Tassis et al. (2012a) (who adopted a ``thin-disk" approximation) however, these simulations usually follow a limited number of chemical species. In our new implementation, we can follow the evolution of up to $\sim$ 300 species, the abundances of which are calculated for every grid cell and timestep. While the inclusion of an extensive number of species is important for accurately calculating the resistivities, it also opens up new possibilities regarding the comparison of such simulations with radio observations of molecular spectra (e.g. Spezzano et al. 2017).

This paper is organized as follows: In section \S~\ref{eqs} we present the basic equations governing the evolution of a three-fluid system with self gravity. In section \S~\ref{resis} we note the expressions for the magnetic resistivities, in \S~\ref{chem} we discuss the chemical networks used from which the resistivities are self-consistently computed and in \S~\ref{numer} we outline the numerical details. In \S~\ref{setup} we describe 2D numerical simulations performed with the \textsc{FLASH} (Fryxell et al. 2000; Dubey et al. 2008) adaptive mesh refinement (AMR) code and in \S~\ref{bench} we compare our results against ideal MHD simulations and previous publications, and present our findings. Finally, we summarize and discuss future prospects in \S~\ref{discuss}.

\section{Methods}
\subsection{Non-ideal MHD effects in collapsing prestellar cores}\label{eqs}

The equations governing the evolution of an isothermal, three-fluid MHD system with self-gravity are:
\begin{subequations}\label{eq:litdiff}

\begin{equation}\label{cntN}
\frac{\partial \rho_n}{\partial t} + \boldsymbol{\nabla} \cdot (\rho_n \boldsymbol{v_n}) = 0
\end{equation}
\begin{eqnarray}\label{momN}
\frac{\partial(\rho_n\boldsymbol{v_n})}{\partial t} + \boldsymbol{\nabla} \cdot (\rho_n \boldsymbol{v_n}\boldsymbol{v_n}) = -\boldsymbol{\nabla}P_n - \rho_n\boldsymbol{\nabla}\psi + \nonumber \\
 \frac{\rho_n}{\uptau_{ni}}(\boldsymbol{v_i} - \boldsymbol{v_n}) + \frac{\rho_n}{\uptau_{ne}}(\boldsymbol{v_e} - \boldsymbol{v_n})
\end{eqnarray}
\begin{equation}\label{momI}
m_in_i\frac{\partial \boldsymbol{v_i}}{\partial t} = en_i(\boldsymbol{E}+\frac{\boldsymbol{v_i}}{c}\times\boldsymbol{B}) + \frac{\rho_i}{\uptau_{in}}(\boldsymbol{v_n} - \boldsymbol{v_i})
\end{equation}
\begin{equation}\label{momE}
m_en_e\frac{\partial \boldsymbol{v_e}}{\partial t} = -en_e(\boldsymbol{E}+\frac{\boldsymbol{v_e}}{c}\times\boldsymbol{B}) + \frac{\rho_e}{\uptau_{en}}(\boldsymbol{v_n} - \boldsymbol{v_e})
\end{equation}
\begin{equation}\label{cur}
\boldsymbol{j} = e(n_i\boldsymbol{v_i} - n_e\boldsymbol{v_e})
\end{equation}
\begin{equation}\label{far}
\frac{\partial \boldsymbol{B}}{\partial t} = -c\boldsymbol{\nabla}\times\boldsymbol{E}
\end{equation}
\begin{equation}\label{amp}
\boldsymbol{\nabla}\times\boldsymbol{B} = \frac{4\pi}{c}\boldsymbol{j}
\end{equation}
\begin{equation}\label{gau}
\boldsymbol{\nabla}^2\psi = 4\pi G\rho_n
\end{equation}
\begin{equation}\label{eos}
P_n = c_s^2\rho_n
\end{equation}
\end{subequations}
In Eqs.~\ref{cntN}---\ref{eos}, $\rho_s$, $n_s$, $m_s$ and $\boldsymbol{v_s}$ denote the mass density, number density, mass, and velocity of species $\textit{s}$, respectively, where $\textit{s}$ = n, i, e for neutrals, molecular and atomic ions, and electrons, treated as the three fluids. The quantities $\boldsymbol{j}$, $P_n$, $\psi$, $\textit{G}$ and $c_s$ denote the total current density, gas pressure, gravitational potential, gravitational constant, and sound speed, respectively. The electric and magnetic field are denoted with $\boldsymbol{E}$, $\boldsymbol{B}$. On the right-hand side of Eqs.~\ref{momN},~\ref{momI},~\ref{momE} the term $\boldsymbol{F}_{sn}=-\boldsymbol{F}_{ns}=\frac{\rho_s}{\uptau_{sn}}(\boldsymbol{v_n}-\boldsymbol{v_s})$ is the frictional force per unit volume on species $\textit{s}$ due to elastic collisions with $\rm{H_2}$ (Ciolek \& Mouschovias 1993; Tassis \& Mouschovias 2005). Here, $\uptau_{sn}$ is the mean collision timescale between species $\textit{s}$ and $\rm{H_2}$, given by:
\begin{equation}\label{mct}
\uptau_{sn} = \frac{1}{\alpha_{s\rm{He}}}\frac{m_s+m_{\rm{H_2}}}{\rho_{\rm{H_2}}}\frac{1}{\mean{\sigma w}_{s\rm{H_2}}}
\end{equation}
(Mouschovias 1995) where $\mean{\sigma w}_{sH_2}$ is the mean collision rate between species $\textit{s}$ and $\rm{H_2}$, and $\alpha_{sHe}$ is a factor to account for the slowing-down time of species  $\textit{s}$ due to collisions with $\rm{He}$. For electrons and ions $\alpha_{sHe}$ = 1.16, 1.14 respectively (Mouschovias 1995). The mean collision rate for ion-neutral collisions is computed from the Langevin approximation (Gioumousis \& Stevenson 1958) and is equal to $\mean{\sigma w}_{iH_2}$ = 1.69$\times$$10^{-9}$ $\rm{cm^{3}}$~$\rm{s^{-1}}$. We use a single value for the mean collision rate for all ions. Pinto et al. (2008) found that the Langevin approximation is justified for collisions between $\rm{H_2}$ and molecular ions for low temperatures, relevant to molecular clouds, and small drift velocities. In Appendix~\ref{vdrift} we show that the maximum ion-neutral drift velocity in the models presented in this study is never more than $0.7\times c_s$ and therefore, both these conditions are satisfied. Compared to the more detailed calculations performed by Pinto et al. (2008), the value of the mean collision rate used here is accurate within 15\% for collisions of $\rm{H_2}$ with $\rm{H_3^+}$ and $\rm{HCO^+}$. These species were found to be the dominant ions in previous studies (Tassis et al. 2012a). For electrons, for which the Langevin approximation does not apply, we have that $\mean{\sigma w}_{eH_2}$ = 1.3$\times$$10^{-9}$ $\rm{cm^{3}}$~$\rm{s^{-1}}$ (Motte \& Massey 1971).

The acceleration terms on the left-hand sides of Eqs.~\ref{momI},~\ref{momE} can be ignored since these species rapidly reach their terminal velocities (see Fig. 2 in Mouschovias et al. 1985). Furthermore, we have ignored thermal and gravitational forces in ~\ref{momI},~\ref{momE} since they are negligible in comparison to the Lorentz force and the forces due to collisions with neutrals. Finally, we have omitted all other species other than neutrals in Eqs.~\ref{gau},~\ref{eos} since, for molecular-cloud conditions, neutrals will always be far more abundant than charged species.

Adding Eqs.~\ref{momI},~\ref{momE}, using Eqs.~\ref{cur} \&~\ref{amp} and the charge neutrality condition ($n_e$ = $n_i$), and noting again that $\boldsymbol{F}_{sn}$=-$\boldsymbol{F}_{ns}$, it can be shown that the last two terms in Eq.~\ref{momN} can be substituted with 1/$4\pi$($\boldsymbol{\nabla}\times\boldsymbol{B})$$\times\boldsymbol{B}$ such that Eq.~\ref{momN} can be re-written as:
\begin{eqnarray}\label{momNnew}
\frac{\partial(\rho\boldsymbol{v_n})}{\partial t} + \boldsymbol{\nabla} \cdot (\rho_n \boldsymbol{v_n}\boldsymbol{v_n}) = -\boldsymbol{\nabla}P_n - \rho_n\boldsymbol{\nabla}\psi \nonumber \\ + \frac{1}{4\pi}(\boldsymbol{\nabla}\times\boldsymbol{B})\times\boldsymbol{B}
\end{eqnarray}
Therefore, even though neutral species do not directly experience the Lorentz force the magnetic field appears in their momentum equation through collisions with charged species.

The advection and evolution of the magnetic field in partially ionized plasmas depends on the conductivity of the fluid. When the fluid has finite conductivity, the current generated due to the electric field which is in turn caused by temporal changes of the magnetic field (see Eq.~\ref{far}), is described by the generalized Ohm's law:
\begin{equation}\label{gohm}
\boldsymbol{E_n} = \boldsymbol{E} + \frac{\boldsymbol{v_n}}{c}\times\boldsymbol{B} = \eta_\perp \boldsymbol{j_\perp} + \eta_\parallel \boldsymbol{j_\parallel} + \eta_H	\boldsymbol{j}\times \boldsymbol{b}
\end{equation}
(see Kunz \& Mouschovias 2009 for a derivation) where $\boldsymbol{E_n}$ is the electric field in the frame of reference of the neutrals, $\eta_\perp$, $\eta_\parallel$ and $\eta_H$ are the perpendicular, parallel and Hall resistivities respectively, $\boldsymbol{b}$ is the unit vector of the magnetic field and $\boldsymbol{j_\perp}$, $\boldsymbol{j_\parallel}$ are the components of the current perpendicular and parallel to the magnetic field.

By considering the curl of the generalized Ohm's law and using Eq.~\ref{far} one can arrive at the magnetic induction equation:
\begin{equation}\label{induc}
\frac{\partial\boldsymbol{B}}{\partial t} = \boldsymbol{\nabla}\times(\boldsymbol{v_n}\times\boldsymbol{B}) - c\boldsymbol{\nabla}\times(\eta_\perp \boldsymbol{j_\perp} + \eta_\parallel \boldsymbol{j_\parallel} + \eta_H	\boldsymbol{j}\times \boldsymbol{b}).
\end{equation}
The physical meaning of Eq.~\ref{induc} is that the magnetic field can be advected with the velocity of the neutrals (as in ideal MHD) with additional terms describing both diffusion due to imperfect coupling through collisions with charged species and dissipation. A much more physical formalism for the induction equation is derived in Tassis \& Mouschovias (2007a) (see their Eq.~20) where the magnetic field is advected with the conductivity-weighted plasma velocity of charged species and additional terms describe dissipation due to currents. Nevertheless, these two formalisms are completely equivalent (see Appendix A3 in Tassis \& Mouschovias 2007a).

Given the structure and already-available routines in the publicly-available version of \textsc{FLASH} we adopt the one-fluid formalism described by Eq.~\ref{induc} with the resistivities computed from our chemical networks, as described below.

\begin{figure*}
\includegraphics[width=2.1\columnwidth, clip]{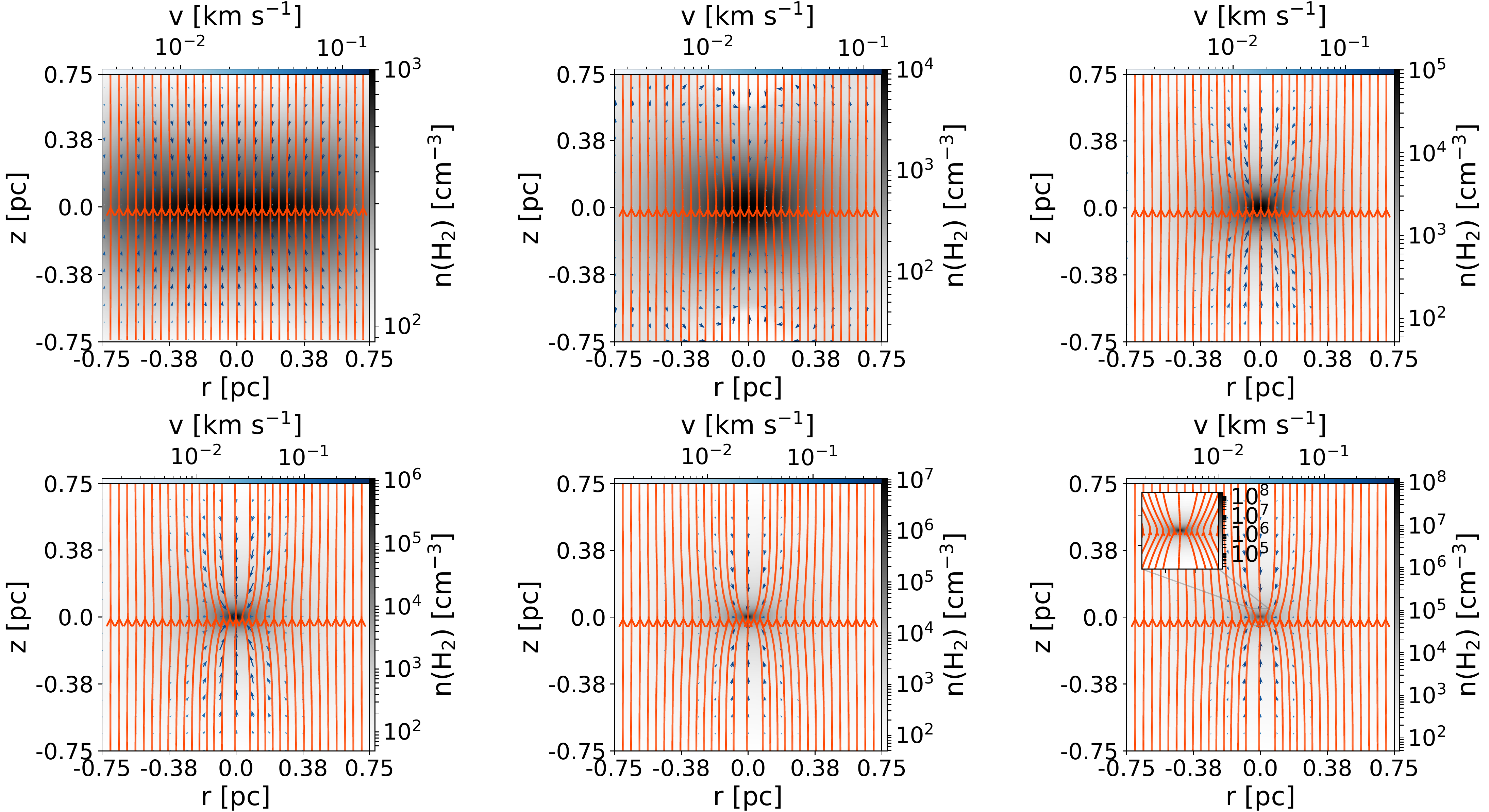}
\caption{Snapshots from simulation \texttt{nI0.5\_noDC\_s$\zeta$\_noG} showing the temporal and spatial evolution of the core. The gray colormap shows the number density of the core, the blue colormap corresponds to the arrows showing the velocity field, and the orange lines show magnetic field lines. From upper left to bottom right the central density is $\sim$$10^3$, $10^4$, $10^5$, $10^6$, $10^7$, $10^8$ $\rm{cm^{-3}}$. In the bottom right panel we additionally show an inset figure where we have zoomed in the inner $\sim$0.1 pc of the core. Given the axisymmetry assumed in our simulations, the $-\rm{R}\leq r<$ 0 portion of these snapshots is the mirrored counterpart of the $0\leq r\leq$ R part of the cloud.
\label{CoreTimeEvolImg}}
\end{figure*}

\subsection{Resistivities}\label{resis}
The expressions for computing the three resistivities that appear in Eq.~\ref{gohm} are:
\begin{gather} 
\eta_\parallel = \frac{1}{\sigma_\parallel},
\qquad
\nonumber \eta_\perp = \frac{\sigma_\perp}{\sigma_\perp^2 + \sigma_H^2},
\qquad \\
\eta_H = \frac{\sigma_H}{\sigma_\perp^2 + \sigma_H^2}
\label{etas}
\end{gather} 
where $\sigma_\parallel$, $\sigma_\perp$ and $\sigma_H$ are the parallel, perpendicular and Hall conductivities. In turn, these three conductivities are defined as:
\begin{gather} 
\sigma_\parallel = \sum_s\sigma_s, \qquad \nonumber \sigma_\perp = \sum_s\frac{\sigma_s}{1+(\omega_s\uptau_{sn})^2}, \\
\sigma_H = -\sum_s\frac{\sigma_s\omega_s\uptau_{sn}}{1+(\omega_s\uptau_{sn})^2}
\label{sigmas}
\end{gather} 
(see Parks 1991). We note here that the units of the resistivities as defined in Eqs.~\ref{etas} are $\rm{s}$, in contrast to Schober et al. (2012) who have absorbed the factor $c/4\pi$ in the definition of the resistivities (see their Eqs 8a \& 8b). In Eqs.~\ref{sigmas}, $\sigma_s$ is the conductivity of species $\textit{s}$ defined as:
\begin{equation}\label{sigs}
\sigma_s = \frac{n_se^2\uptau_{sn}}{m_s}
\end{equation}
In Eqs.~\ref{sigmas} \&~\ref{sigs}, $\omega_s$ is the gyrofrequency of species $\textit{s}$ defined as $\omega_s$ = $q_sB/m_sc$, which also carries the algebraic sign of the charge $q_s$ and $\uptau_{sn}$ is defined in Eq.~\ref{mct}. The number densities of charged species $n_s$ in Eq.~\ref{sigs} are computed from the chemical networks described in \S~\ref{chem} below.

The ambipolar diffusion and Ohmic resistivities can then be computed from the perpendicular and parallel conductivities as:
\begin{gather} 
\eta_{\parallel AD} = 0, \quad \nonumber \eta_{\perp AD} = \eta_{\perp}-\eta_{\parallel}, \\
\eta_{\parallel OD} = \eta_{\parallel}, \quad \eta_{\perp OD} = \eta_{\parallel}
\label{etaadOhm}
\end{gather}
(see Desch \& Mouschovias 2001 and references therein).


\subsubsection{Grains}

Eqs.~\ref{cntN}---\ref{eos} are written without grains taken into account. However, charged grains can also couple to the magnetic field and increase the ambipolar-diffusion timescale (Tassis \& Mouschovias 2004) by slowing down neutral species through collisions (Baker 1979; Elmegreen 1979; Umebayashi \& Nakano 1990; Ciolek \& Mouschovias 1994). The six-fluid MHD equations including neutrals, negative and positive grains can be found in Kunz \& Mouschovias (2009). Here we ignore neutral and positive grains since for densities relevant for prestellar cores (which we follow here - see \S~\ref{bench})) the majority of the grains are negatively charged (Tassis et al. 2012a).

In our implementation we assume either a uniform or a MRN distribution (Mathis et al. 1977) of spherical grains following Kunz \& Mouschovias (2009). In summary, for the uniform distribution we use a grain radius of $\alpha_0$ = 0.0375 $\rm{\upmu m}$ and for the MRN distribution the minimum and maximum radii are $\alpha_{min}$ = 0.0181 $\rm{\upmu m}$ and $\alpha_{max}$ = 0.9049 $\rm{\upmu m}$, respectively. In contrast to Marchand et al.(2016) who use ten bins of grain radii we adopt thirty bins. In both the uniform and MRN distribution case, the mass density of the grains is $\rm{\uprho_{g^-}}$ = 2.3 g $\rm{cm^{-3}}$. Finally, we assume a fixed abundance of $n_{g^-}/n_{\rm{H_2}} = 10^{-12}$ which is treated as a free parameter (Tassis et al. 2012a). 

Given that we assume a fixed grain abundance (i.e. no charge transfer between the grains and gas), the polarization factor for grains (Draine \& Sutin 1987) is neglected. However, adopting a fixed grain abundance is a reasonable assumption for the density range studied here. Previous calculations where charge transfer between the grains and the gas has been taken into account (see e.g. Fig. 6 from Kunz \& Mouschovias 2010) have found that the grain abundance remains roughly constant for number densities $n_{\rm{H_2}}\le10^9~\rm{cm^{-3}}$. A higher dust abundance would lead to a better coupling between the neutrals and the magnetic field and a cloud would take longer to collapse. On the other hand, the freeze out of species onto dust grains in our chemical model (see \S~\ref{chem}) depends on the number density of grains. Therefore, a higher dust abundance would lead to more depletion, ambipolar diffusion would proceed faster and thus the cloud would collapse faster.

When grains are taken into account (see \S~\ref{sims}) the charge-neutrality condition becomes:
\begin{equation}\label{chargeneutralGrains}
n_i = n_e + n_{g^-}
\end{equation}
and the grains contribute to the resistivities as described in Eqs.~\ref{mct} \& \ref{etas}--\ref{sigs}. The mean collision rate for the grains in Eq.~\ref{mct} is calculated as:
\begin{equation}\label{mcrgrains}
\mean{\sigma w}_{g^-H_2} = \pi \alpha^2(8k_BT/\pi m_{H_2})^{1/2}
\end{equation}
(Ciolek \& Mouschovias 1993) where $k_B$ is the Boltzmann constant. The corresponding $\alpha_{sHe}$ factor in Eq.~\ref{mct} for grains is equal to 1.28 (Tassis \& Mouschovias 2005). Conversely, when grains are assumed to be neutral, they are not taken into account in the calculation of resistivities but they still contribute to the chemistry of the cloud through the processes described in the next section.


\subsection{Chemical model}\label{chem}

We have currently implemented two chemical networks in \textsc{FLASH}. The large chemical networks consists of $\sim$ 300 species, 82 of which are in the dust phase. The evolution of these species is determined by $\sim$14$\times$$10^{3}$ chemical reactions. This chemical network was first presented in Tritsis et al. (2016) (see their Table 1 for a list of all the species included in the network). The small chemical networks does not include any deuterium chemistry but is otherwise identical to our large network. In total, it consists of 78 gas-phase species and 37 grain species and $\sim$1650 chemical reactions. The omission of deuterium chemistry leads to a factor of $\sim$40--140 increase in computation speed. In turn, each timestep in dynamical models without any non-equilibrium chemistry included is on average a factor of $\sim$230 less expensive than the models where we include our small network.

\begin{table}
\begin{center}
\begin{tabular}{l r}
\hline\hline
 Species & Abundance \\ 
 \hline
 H                & 1.55$\times$$10^{-2}$ \\  
 He               & 2.17$\times$$10^{-1}$ \\
 C$^{+}$          & 1.13$\times$$10^{-4}$ \\
 N                & 3.32$\times$$10^{-5}$ \\
 O                & 2.73$\times$$10^{-4}$ \\
 Si$^{+}$         & 3.10$\times$$10^{-8}$ \\
 $\rm{H_2}$       & 7.67$\times$$10^{-1}$ \\
 $\rm{D_2}$       & 6.18$\times$$10^{-6}$ \\
 D                & 2.48$\times$$10^{-7}$ \\
 HD               & 1.23$\times$$10^{-5}$ \\
\hline\hline
\end{tabular}
\end{center}
\caption{\label{initabund} Initial abundances of the species used in our chemical models relative to total hydrogen nuclei.}
\end{table}

\begin{figure*}
\includegraphics[width=2.1\columnwidth, clip]{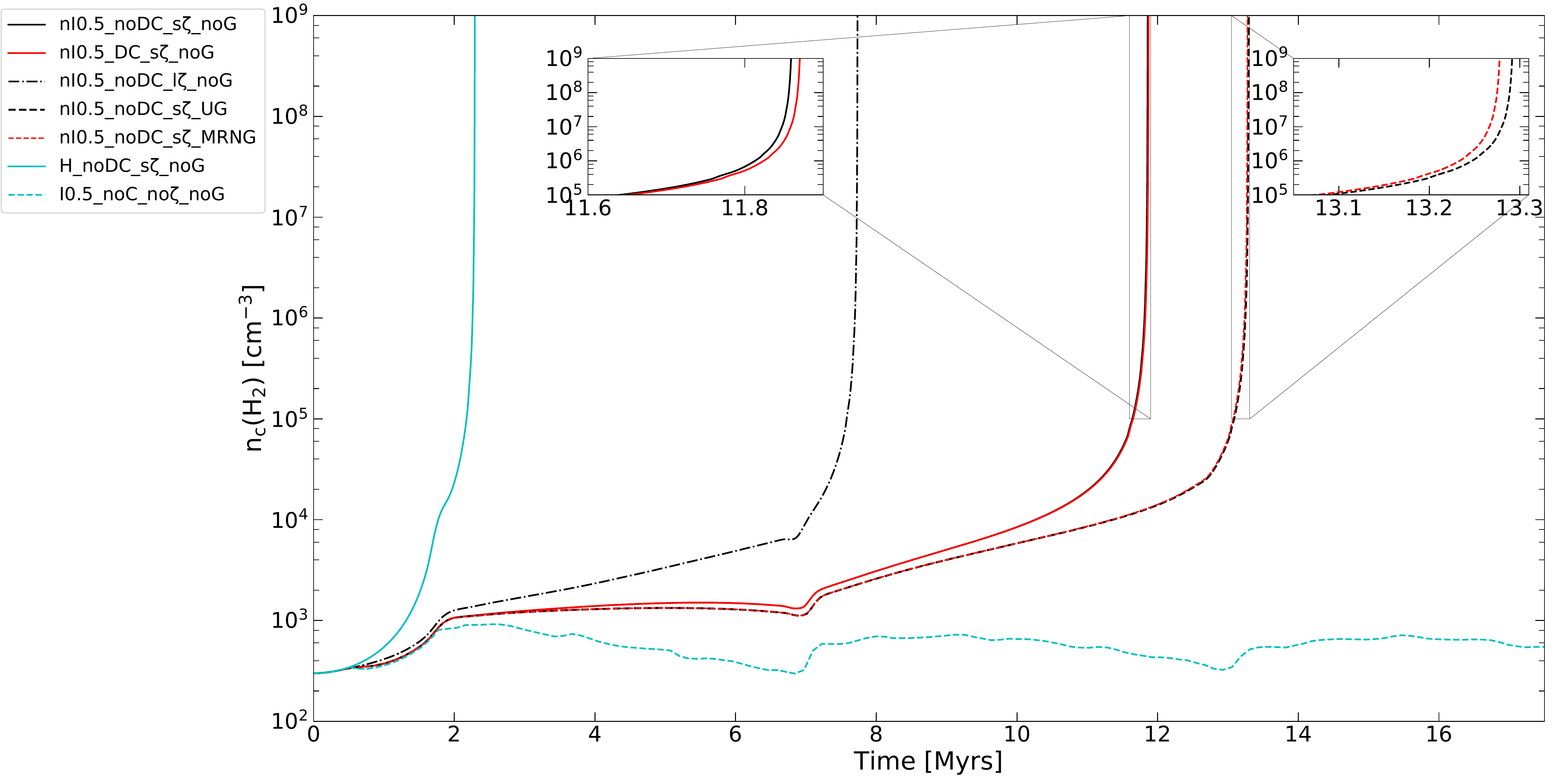}
\caption{Evolution of the central density of the core as a function of time for all the models listed in Table~\ref{simsParams}. In the inset figures we highlight the differences between models \texttt{nI0.5\_noDC\_s$\zeta$\_noG} and \texttt{nI0.5\_DC\_s$\zeta$\_noG}, and \texttt{nI0.5\_noDC\_s$\zeta$\_UG} and \texttt{nI0.5\_noDC\_s$\zeta$\_MRNG}.
\label{CdensTimeEvol}}
\end{figure*}

Reaction-rate coefficients for both networks are taken from the fifth release of the $\textsc{UMIST}$ database for astrochemistry (McElroy et al. 2013). We model the following processes for the grains: \begin{enumerate*}[label=(\roman*)]
\item freeze-out,
\item thermal desorption,
\item photo desorption,
\item cosmic-ray heating and
\item grain-surface chemical reactions.
\end{enumerate*}
Initially, the only species in molecular form is $\rm{H_2}$ and in the case of the large chemical network $\rm{D_2}$ and $\rm{HD}$. In Table~\ref{initabund} we list the initial elemental abundances relative to the total density. Considering these values, the mean molecular weight is $\sim$~2.4, the C/O ratio is $\sim$0.4 and the D/H ratio is 1.6$\times$$10^{-5}$.

The initial elemental abundances listed in Table~\ref{initabund} are, within uncertainties, in agreement with observational estimates. For instance, Fuente et al. (2019) found that a C/O ratio of 0.4 could explain the observed N(HCN)/N(CO) ratio in TMC-1. Likewise, for $\zeta$ Oph Cardelli et al. (1993) found a C/O ratio of $0.45\pm0.11$. Linsky (2003) discussed a number of observations and found that the D/H ratio in the Local Bubble is 1.5$\times$$10^{-5}$. We additionally note here that the results from our large network where found to be in agreement with those by Tassis et al. (2012a) who in turn compared their results with a number of observational studies (Tritsis et al. 2016).

For the simulations presented in \S~\ref{sims} we use a constant $\rm{A_v}$ of 10 which is meant to represent a prestellar core embedded in a molecular cloud. We additionally use a constant cosmic-ray ionization rate. A relation that is often used to describe the scaling of the cosmic-ray ionization rate with column density is:
\begin{equation}
\zeta_{cr} = \zeta_0~\rm{exp}(-\Sigma_{H_2}/96~g~cm^{-2})
\end{equation}
(Umebayashi 1983) where $\zeta_0 = 1.3\times10^{-17} \rm{s^{-1}}$ is the standard value for the cosmic-ray ionization rate (Spitzer 1978; Caselli et al. 1998; Bovino et al. 2020) and $\rm{\Sigma_{H_2}}$ is the mass column density of $\rm{H_2}$. Following this relation, one expects a significant decrease in the cosmic-ray ionization rate for number densities $n_{\rm{H_2}} \gg 10^{10} \rm{cm^{-3}}$. However, since we do not follow the evolution of our simulated cores past these densities (see \S~\ref{sims}), a constant cosmic-ray ionization rate is sufficient.

The significance of adding non-equilibrium chemical modeling to dynamical simulations was thoroughly discussed in Tassis et al. (2012a) with whom we employ similar methods in terms of the chemical modeling. Specifically, Tassis et al. (2012a) examined the ratio of the total production to destruction rates and found that several species in their chemical network remained out of equilibrium throughout the evolution of their models (see their Fig. 3 and the discussion in their section 5.2). Similarly to Tassis et al. (2012a), several species in our chemical networks remain out of equilibrium and their number densities change significantly even if dynamically, our simulations are in quasi-equilibrium. This is due to the fact that the timescale needed to achieve chemical steady state is multiple times the timescale of the slowest chemical reaction. Therefore, the steady-state chemical timescale is always much longer than the dynamical timescale and the network remains out of equilibrium.


\subsection{Numerical implementation}\label{numer}

\subsubsection{Chemistry}\label{chemnum}

Each chemical species in our chemical networks is treated as a different fluid. The chemical species are treated as passive scalars and are advected with the velocity of the neutrals. Therefore, the advection equation of species \textit{s} is given by:
\begin{equation}\label{cntNspecies}
\frac{\partial \rho_s}{\partial t} + \boldsymbol{\nabla} \cdot (\rho_s \boldsymbol{v_n}) = S_j
\end{equation}
where $\rho_s$ is the mass density of species \textit{s}, and $S_j$ are source/sink terms because of chemical reactions. The abundances of the species due to source/sink terms are calculated after advection terms by solving a system of ordinary differential equations described by our chemical network. To solve the system of ordinary differential equations we use the Livermore Solver for Ordinary Differential Equations (Hindmarsh 2019). The chemical timestep is the same as the hydrodynamical timestep as this is computed from the modified Courant--Friedrichs--Lewy condition (Courant et al. 1928; see \S~\ref{modCourant}). The abundances of $\rm{H_2}$ and \textit{e} are calculated from the conservation of total hydrogen and the charge-neutrality condition respectively. Likewise, for our large chemical network, the abundance of $\rm{D_2}$ is calculated from the conservation of the total deuterium.

The approximation that all chemical species (including ions) are advected with the velocity of the neutrals is justified on the basis that the characteristic (i.e. shortest) timescale for the production or destruction of charged species is much shorter than the dynamical timescale. Consequently, any differences that would appear as a result of advecting the species with different velocities would be quickly eliminated due to the chemical reactions. For instance, in the simulations presented herewith (see \S~\ref{sims}), the chemical timescale for the production of $\rm{H_3^+}$ and $\rm{D_3^+}$, at central number densities of $n_c\sim10^3$--$10^5~\rm{cm^{-3}}$ (where a subcritical cloud spends the majority of its lifetime), is $\le10^{-3}$ yrs in the inner regions of the cloud. For the same central number densities, the chemical timescale in the outer regions of the cloud is typically 0.01--0.5 yrs. In comparison, the dynamical timescale ranges from a few years in the outer regions of the cloud to a few thousand years in the inner regions of the cloud. Therefore, this approximation is well justified and always holds true in the inner parts of the cloud which are of interest.


\subsubsection{Currents}

To compute the components of the current in the perpendicular and parallel directions to the magnetic field we simply compute the current from Eq.~\ref{far} and then perform a vector projection as:
\begin{gather} 
\boldsymbol{j_\parallel} = \boldsymbol{j}\cdot\frac{\boldsymbol{B}}{\norm{\boldsymbol{B}}}, \qquad
\boldsymbol{j_\perp} = \boldsymbol{j} - \boldsymbol{j_\parallel}
\label{perpCur}
\end{gather}
For axially-symmetric simulations where there is no toroidal component of the magnetic field ($\boldsymbol{j_\parallel}$=0), Eq.~\ref{induc} becomes:
\begin{equation}\label{induc2D}
\frac{\partial\boldsymbol{B}}{\partial t} = \boldsymbol{\nabla}\times(\boldsymbol{v_n}\times\boldsymbol{B}) - \boldsymbol{\nabla}\times(\frac{c^2\eta_\perp}{4\pi}\boldsymbol{\nabla}\times\boldsymbol{B})
\end{equation}
and no vector projection is necessary.


\begin{table*}
\begin{center}
\begin{tabular*}{\textwidth}{c @{\extracolsep{\fill}} c c c c c c c}
\hline\hline
& Model name & Chemistry & $\zeta$/$\zeta_0$ & $\rm{\mu_{ref}}$ & non-ideal MHD & Grains & Visual \\ 
 \hline
Fiducial        & \texttt{nI0.5\_noDC\_s$\zeta$\_noG} & without D & 1 & 0.5 & \checkmark & \xmark & \includegraphics[scale=0.5]{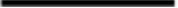} \\
                & \texttt{nI0.5\_DC\_s$\zeta$\_noG} & with D & 1 & 0.5 & \checkmark & \xmark & \includegraphics[scale=0.5]{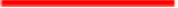} \\
Ionization rate & \texttt{nI0.5\_noDC\_l$\zeta$\_noG} & without D & 0.1 & 0.5 & \checkmark & \xmark & \includegraphics[scale=0.5]{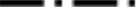} \\
Grains          & \texttt{nI0.5\_noDC\_s$\zeta$\_uG} & without D & 1 & 0.5 & \checkmark & uniform & \includegraphics[scale=0.5]{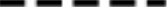} \\
                & \texttt{nI0.5\_noDC\_s$\zeta$\_MRNG} & without D & 1 & 0.5 & \checkmark & MRN & \includegraphics[scale=0.5]{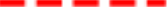} \\
Hydro           & \texttt{H\_noDC\_s$\zeta$\_noG} & without D & 1 & \xmark & \xmark & \xmark & \includegraphics[scale=0.5]{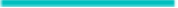} \\
Ideal MHD       & \texttt{I0.5\_noC\_no$\zeta$\_noG} & \xmark & \xmark & 0.5 & \xmark & \xmark & \includegraphics[scale=0.5]{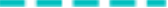} \\
Convergence & \texttt{nI0.5\_noDC\_s$\zeta$\_noG\_conv} & without D & 1 & 0.5 & \checkmark & \xmark & \includegraphics[scale=0.5]{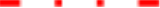} \\
\hline\hline
\end{tabular*}
\end{center}
\caption{\label{simsParams} List of the simulations performed. In naming our models, ``\texttt{nI}", ``\texttt{I}" and ``\texttt{H}" refer to non-ideal, ideal and non-magnetic (hydrodynamical) simulations. What follows is the value of the mass-to-flux ratio, whether we use the network that includes deuterium chemistry or the one that does not (``\texttt{DC}" and ``\texttt{noDC}" respectively), whether we use the standard or a lower value for the cosmic-ray ionization rate (``\texttt{s$\zeta$}" and ``\texttt{l$\zeta$}" respectively) and finally whether we include grains in computing the resistivities or not (``\texttt{uG}"/``\texttt{MRNG}" for a uniform and a MRN distribution (Mathis et al. 1977) respectively and ``\texttt{noG}" when the grains are assumed to be neutral). We stress here that ``noG" only refers to whether we use grains for computing the resistivities and \textit{not} for the chemical modeling. In the last column we list the color/line-style we use throughout this study (unless the meaning of line-styles/colors is otherwise explicitly stated) to show the results from each model.}
\end{table*}

\subsubsection{Timestep}\label{modCourant}

We have also modified the \textsc{Diffuse} unit of \textsc{FLASH} such that the timestep is computed as:
\begin{equation}\label{cfl}
\Delta t \le \Delta t_{\rm{diff}} = \frac{4\pi}{c^2\eta_{\perp}}\frac{min(\Delta x^2,~\Delta y^2,~\Delta z^2)}{2}
\end{equation}
(Mac Low et al. 1995) where $\Delta x$, $\Delta y$ and $\Delta z$ are the minimum sizes of all grid cells across all computational blocks, in the \textit{x}, \textit{y}, and \textit{z} directions, respectively. Therefore, the timestep in the simulations presented here (see \S~\ref{sims}) can be several orders of magnitude smaller compared to the ideal MHD case. Consequently, for a significant time in the evolution of some models presented in \S~\ref{sims} the timestep is just $\sim$2 yrs. As a result, more that 1.9 million timesteps and $\sim10^6$ CPU hours were required for some of these models to reach a central density of $10^9$~$\rm{cm^{-3}}$.

The small timestep computed from Eq.~\ref{cfl} and quoted above, might at first seem contradictory with the approximation discussed in \S~\ref{chemnum}. However, such small timesteps are primarily driven by the outer, low-density parts of the cloud where the resistivity can become artificially large. In the context of the approximation discussed in \S~\ref{chemnum}, this small timestep only poses an issue in a very thin layer close to the z-boundaries of our simulations (where $n_{\rm{H_2}}\le100~\rm{cm^{-3}}$; see \S~\ref{sims}) for central number densities $n_c\ge10^7~\rm{cm^{-3}}$. Even so, we do not set the resistivities to zero or to a constant value in this thin layer, to avoid imposing a ``floor".

\section{Simulations}\label{sims}                                                                             %

\subsection{Numerical setup}\label{setup}

We use the unsplit staggered mesh algorithm (Lee 2013; Lee \& Deane 2009) to perform a suite of 2D axially-symmetric numerical simulations in cylindrical coordinates. Here, we summarize the initial and boundary conditions adopted. The initial number density in all of our simulations is 300~$\rm{cm^{-3}}$ and the magnetic field for all magnetic models is aligned with the $z$-axis (i.e. the axis of symmetry). In all our simulations the initial mass-to-flux ratio is 0.5 (in units of the critical value) and the strength of the magnetic field is 15~${\upmu}$G. We use an isothermal equation of state with the temperature of the cloud set to $\rm{T} = 10~K$. Finally, all the velocity components are initially set to zero.

\begin{figure}
\includegraphics[width=1.0\columnwidth, clip]{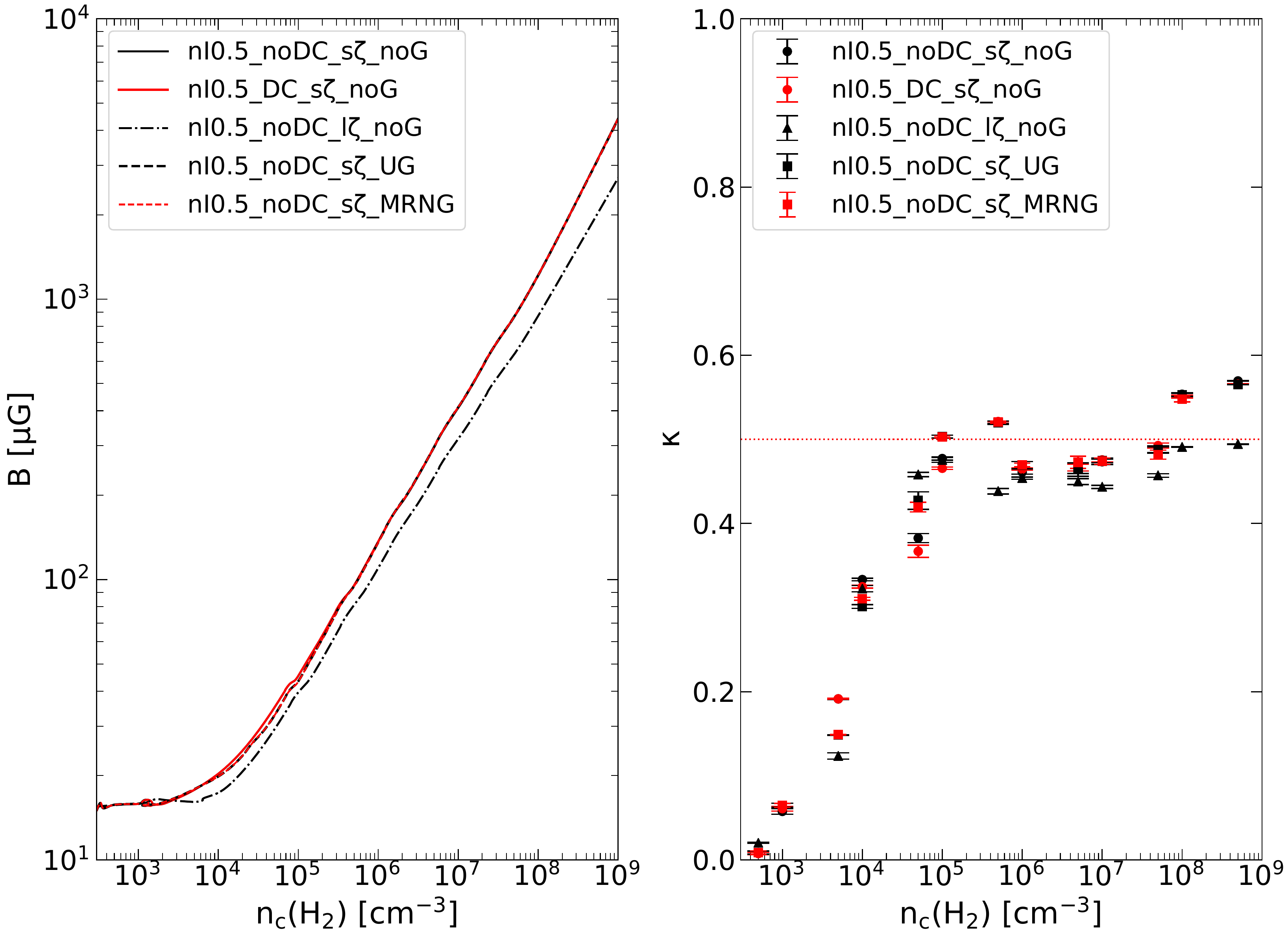}
\caption{Left panel: evolution of the magnetic field as a function of density from our simulations. Right panel: the value of the exponent $\upkappa$ from the magnetic field-gas density relation as a function of density. The value of the exponent $\upkappa$ is measured in intervals of half an order of magnitude. The red dotted line marks the theoretically expected value of 0.5 (see Tritsis et al. 2015 and references therein).
\label{Brho}}
\end{figure}

Given the axisymmetry assumed, and in order to reduce computational cost, we only follow the evolution of half the cloud (i.e. $0\leq r\leq$ R, $-Z\leq z\leq Z$). The radius $\textit{R}$ and half-thickness of the cloud $\textit{Z}$ are both set equal to 0.75 pc. The boundary condition at $r=R$ is set to diode. For $z=-Z$ and $z=Z$ we use periodic boundaries conditions so that we do not lose any mass along magnetic field lines and artificially alter the mass-to-flux ratio as the simulation progresses. Finally, we use an axisymmetric boundary condition at $r=0$. We use an adaptive mesh grid with an initial size of $64\times128$ cells and six levels of refinement (including the first AMR level). The physical size of the finest grid cell is therefore $\sim$76 au. The AMR refinement is based on the modified second derivative criterion of the density and $z$-component of the magnetic field, as described in L{\"o}hner (1987). We use the multipole solver (Couch et al. 2013) for Poisson's equation, the HLL Riemann solver (Einfeldt et al. 1991) and the van Leer slope limiter (van Leer 1974). Given that we employ the unsplit staggered mesh algorithm that uses the constrained transport method (Evans \& Hawley 1988), the condition that $\rm{\boldsymbol{\nabla}\cdot\boldsymbol{\rm{B}}} = 0$ is satisfied at all times to machine precision.

In Table~\ref{simsParams} we list the simulations we perform together with the corresponding parameters. In total, we perform a suite of seven simulations and an additional one to check for convergence (see Appendix~\ref{convergence}). We run two simulations with $\upmu$/$\upmu_{crit}=0.5$, considering the standard cosmic-ray ionization rate and each of our chemical networks. We perform another simulation with our small chemical network where we decrease the cosmic-ray ionization rate to 0.1 of its standard value. To test for the effect of grains we additionally run two simulations, again with our small chemical network, where we take grains into account when computing the resistivites and use either a uniform or a MRN distribution. Finally, we run one ideal MHD simulation with $\upmu$/$\upmu_{crit}=0.5$ for benchmarking and to test for numerical diffusion, and one hydrodynamical simulation to explore the chemical abundances in non-magnetic models.

\subsection{Benchmarks \& Results}\label{bench}

\subsubsection{Time Evolution}

In Fig.~\ref{CoreTimeEvolImg} we show six snapshots of the core from model \texttt{nI0.5\_noDC\_s$\zeta$\_noG}. The orange field lines show the magnetic field and the blue colour-coded vectors show the velocity field. Moving from the upper left to the bottom right panel the central density in each snapshot increases by one order of magnitude. Initially, the cloud contracts along magnetic field lines (upper left panel). During this phase, the strength of the magnetic field does not scale with density, i.e. $B\propto\rho^0$ (see also Fig.~\ref{Brho}) and magnetic-field lines remain straight. At the end of this phase which lasts for approximately one free-fall time, the number density is $10^3$ $\rm{cm^{-3}}$. 

The cloud then slightly bounces back as a result of exceeding force balance along magnetic-field lines (its equilibrium state) and enters a phase where it evolves quasi-statically (see also Fig.~\ref{CdensTimeEvol}).  During this phase, the magnetic field does not scale with density (see Fig.~\ref{Brho}) which is however steadily increasing as a result of ambipolar diffusion. At the end of this phase, the number density is $\sim$$10^4$ $\rm{cm^{-3}}$ and the cloud's mass-to-flux ratio has exceeded the critical value.

Finally, the core enters a phase of dynamical contraction which lasts for approximately 1.6 Myrs. For the models where grains are taken into account when computing the resistivities this timescale is increased to 1.9 Myrs whereas for model {nI0.5\_noDC\_l$\zeta$\_noG} where the cosmic-ray ionization rate is 0.1 of the standard value this timescale is reduced to $\sim$ 0.7 Myrs. Such timescales are in agreement with recent estimates of the lifetime of L1512 (Lin et al. 2020). During this phase the magnetic field scales as $B\propto\rho^\upkappa$ where $\upkappa=0.45-0.56$. (see Fig.~\ref{Brho}). The magnetic field is dragged inwards and an hourglass morphology is formed in the innermost parts of the cloud. This is particularly evident in the bottom three panels and especially in the inset of the bottom right panel showing the inner $\sim$0.1 portion of the core. In the outer parts, or the ``envelope" of the cloud, the magnetic field lines remain unperturbed and the radial component of the velocity is nearly zero for radii $\gtrsim$ 0.35 pc. This indicates that only in the innermost magnetic-flux tubes the mass-to-flux ratio exceeds the critical value.

In Fig.~\ref{CdensTimeEvol} we show the evolution of the central density of the core (i.e. at $r=z=0$) as a function of time from all our models. All our magnetic models, excluding the ideal-MHD case, reach approximately the same central density after one free-fall time. In our ideal MHD model (\texttt{I0.5\_noC\_no$\zeta$\_noG}) the value of the central density reached after one free-fall time is $\sim$10\% lower than that reached in the models where non-ideal MHD effects are included. Furthermore, the cloud in this model completely bounces back with the density reaching approximately the same value as in the initial conditions (e.g. 300~$\rm{cm^{-3}}$. The cloud then dynamically oscillates and, without any numerical diffusion, would do so indefinitely. We have run this model for up to 120 Myrs (see Appendix~\ref{convergence}) and numerical diffusion is not sufficient for the cloud's mass-to-flux ratio to exceed the critical value and collapse. Therefore, while it comes as no surprise that a cloud with $\upmu$/$\upmu_{crit}=0.5$ does not collapse under ideal-MHD conditions, this is a useful experiment to test for numerical diffusion.

\begin{figure*}
\includegraphics[width=2.1\columnwidth, clip]{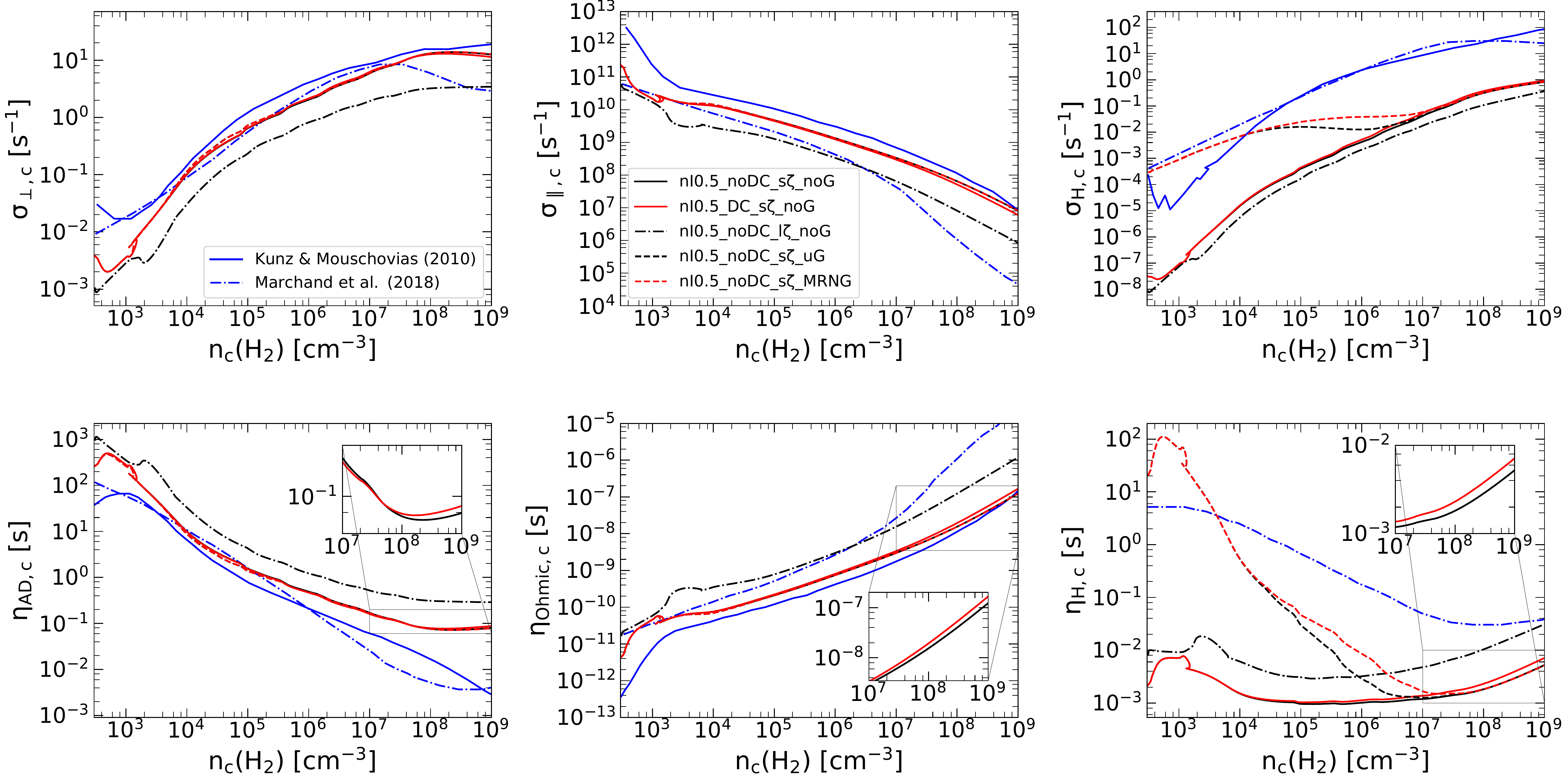}
\caption{Evolution of the perpendicular (upper left panel), parallel (upper middle panel) and Hall (upper right panel) conductivities as a function of central density. The Hall conductivity computed in Marchand et al. (2016) for the number-density range shown here is negative and we therefore show their $-\sigma_H$. From left to right, in the lower row, we show the ambipolar diffusion, Ohmic and Hall resistivities respectively. In the inset figure shown in each panel of the lower row we highlight the differences that are starting to emerge in the corresponding resistivites between models $\texttt{nI0.5\_noDC\_s$\zeta$\_noG}$ and $\texttt{nI0.5\_DC\_s$\zeta$\_noG}$ for densities $n_{\rm{H_2}}\ge 10^7~\rm{cm^{-3}}$. With the blue solid and dashed-dotted lines we show the results by Kunz \& Mouschovias (2010) and Marchand et al. (2016), respectively. The rest of the lines/colors are as described in Table~\ref{simsParams}.
\label{conductResi}}
\end{figure*}

\begin{figure}
\includegraphics[width=1.0\columnwidth, clip]{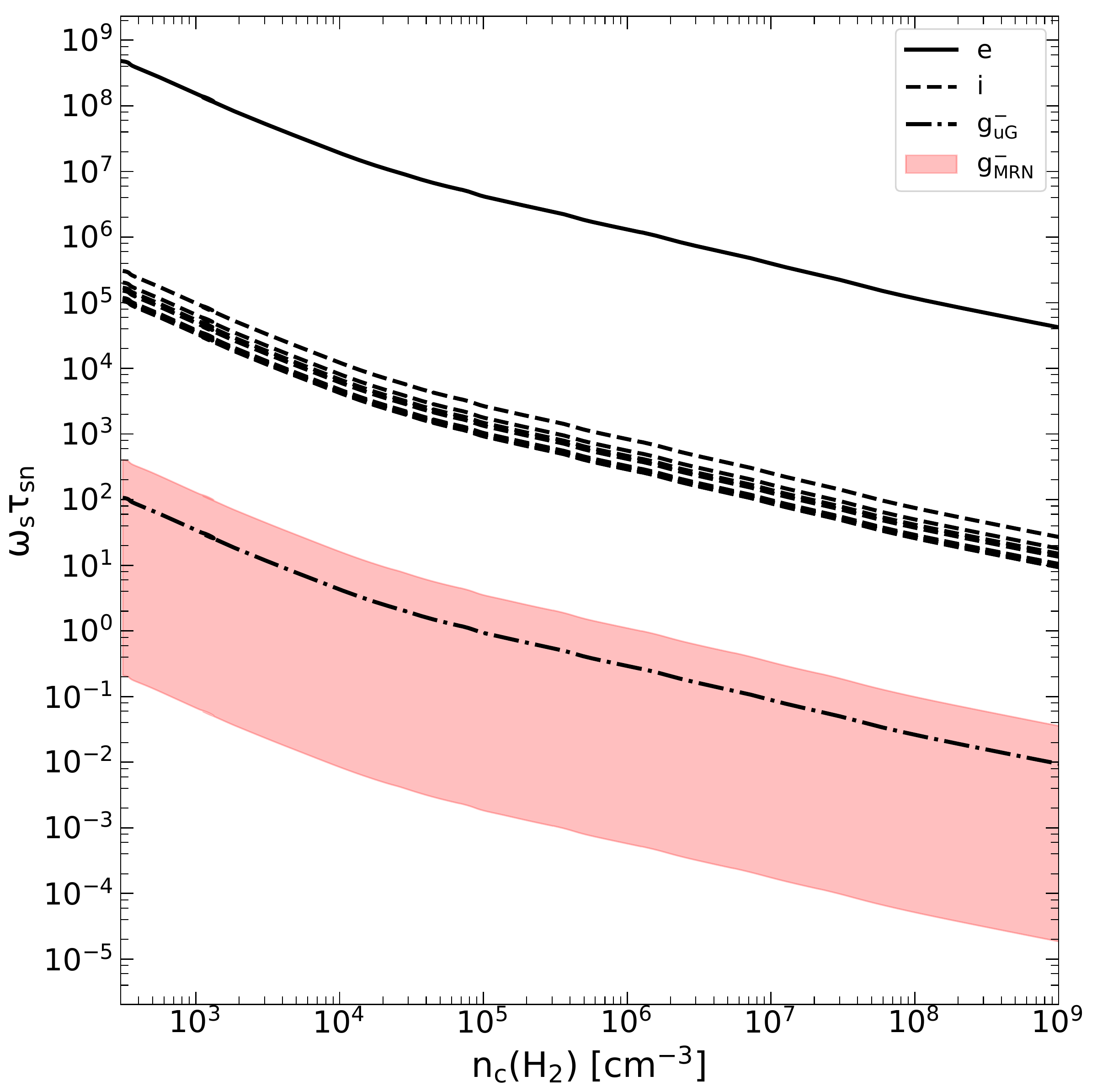}
\caption{The factor $\omega_s\uptau_{sn}$ (see Eq.~\ref{mct} and section \S~\ref{resis}) as a function of the central density of the core. When $\omega_s\uptau_{sn}$~$\gg$~1 species ``\textit{s}" is well-attached to the magnetic field whereas the opposite is true when $\omega_s\uptau_{sn}$~$\ll$~1.
\label{attachFact}}
\end{figure}

\begin{figure*}
\includegraphics[width=2.1\columnwidth, clip]{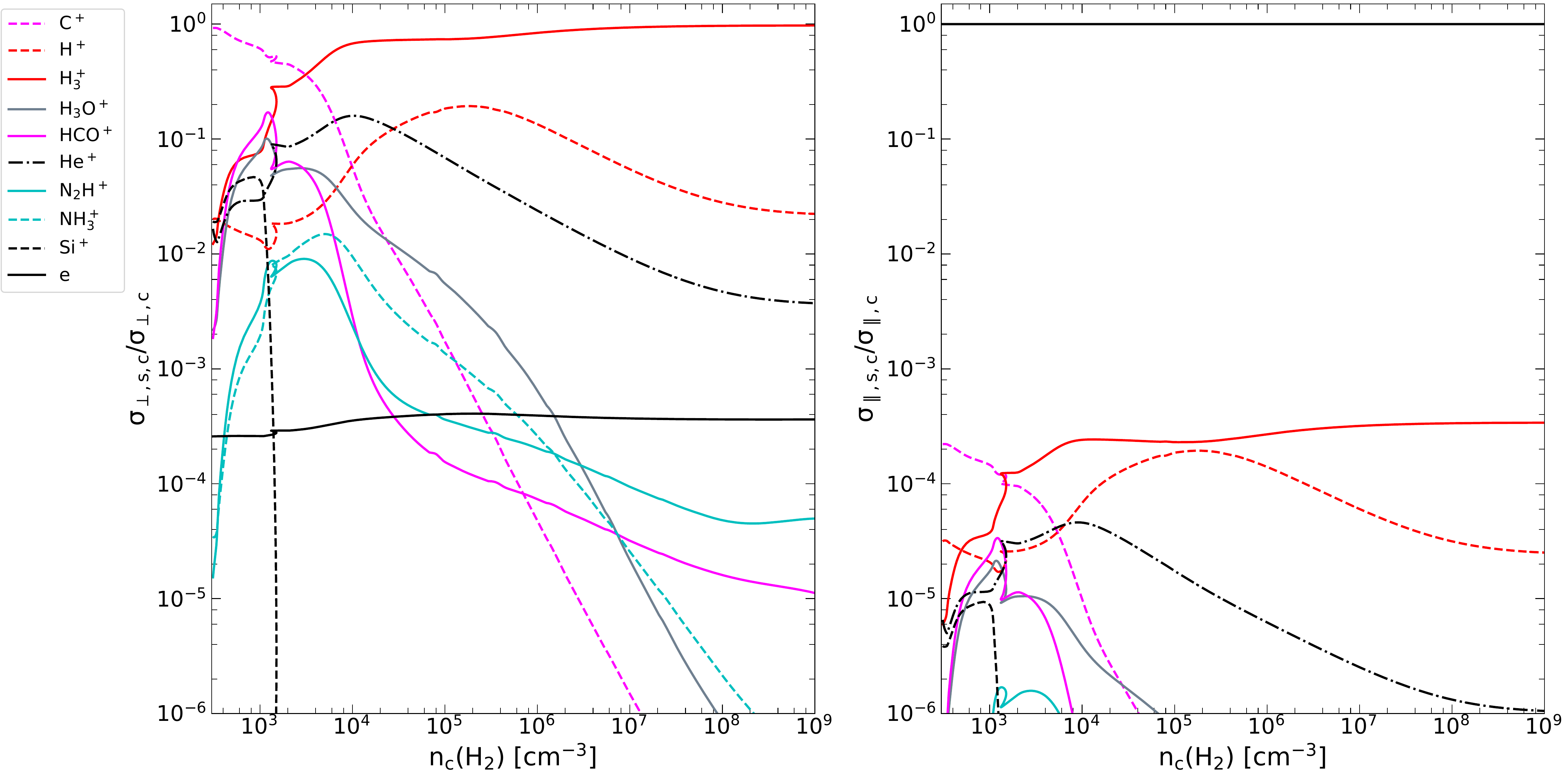}
\caption{The contribution of individual species to the perpendicular (left) and parallel (right) conductivities at the centre of the core as a function of the central $\rm{H_2}$ number density for model $\texttt{nI0.5\_noDC\_s$\zeta$\_noG}$. We only plot the species that, at any given time in the evolution of the core, are in the five most important species that carry either the perpendicular or parallel conductivity. When adopting our small chemical network, we find that the dominant ion and the species that caries most of the perpendicular conductivity is $\rm{H_3^+}$ and not $\rm{HCO^+}$ as was found in earlier studies (e.g. Desch \& Mouschovias 2001; Tassis \& Mouschovias 2007a). This is in agreement with the results by Tassis et al. (2012b).
\label{conductContributions_fidu}}
\end{figure*}

\begin{figure*}
\includegraphics[width=2.1\columnwidth, clip]{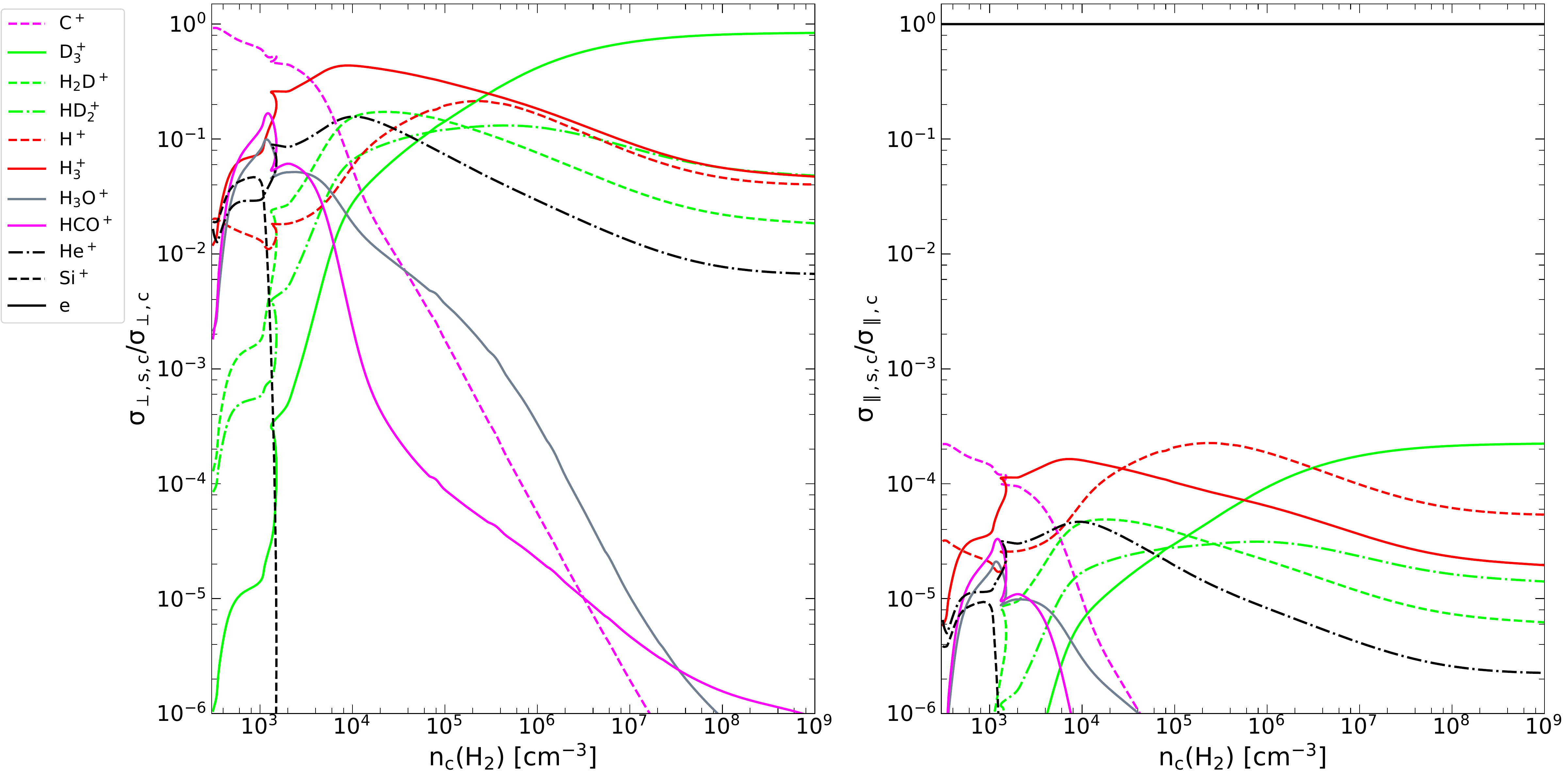}
\caption{Same as in Fig.~\ref{conductContributions_fidu} but for model $\texttt{nI0.5\_DC\_s$\zeta$\_noG}$ where we adopted our large chemical network which includes deuterium chemistry. The situation drastically changes compared to Fig.~\ref{conductContributions_fidu}. While $\rm{H_3^+}$ remains the dominant species for the early stages of the evolution, at higher densities, the ion that carries most of the perpendicular and parallel conductivities is $\rm{D_3^+}$ instead of $\rm{H_3^+}$. Furthermore, at densities beyond $10^7~\rm{cm^{-3}}$ $\rm{H_2D^+}$ becomes as important as $\rm{H_3^+}$. Finally, the relative importance of $\rm{H_3^+}$ compared to $\rm{H^+}$ is also reduced.
\label{conductContributions_DC}}
\end{figure*}


\begin{figure*}
\includegraphics[width=2.1\columnwidth, clip]{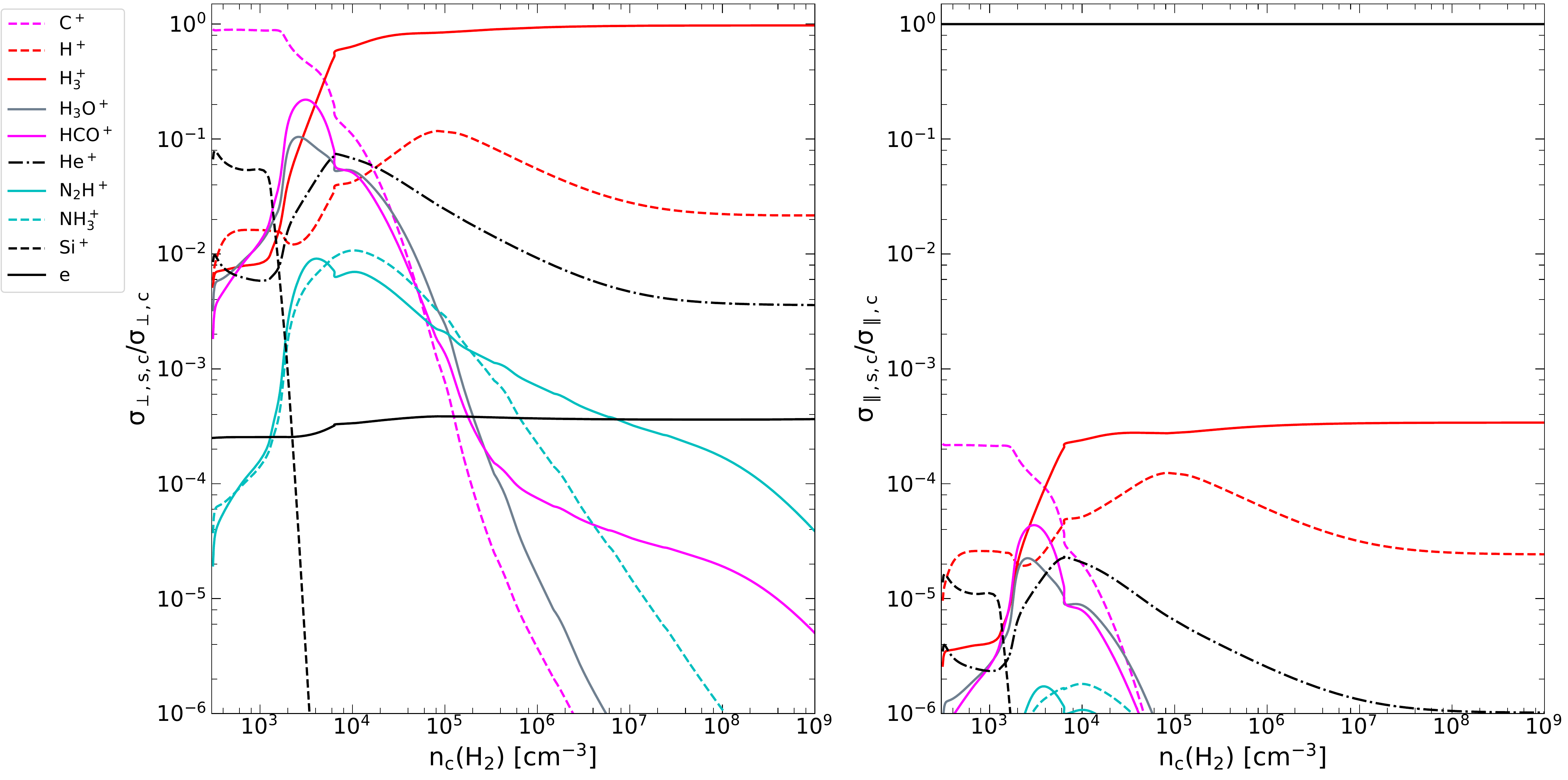}
\caption{Same as in Fig.~\ref{conductContributions_fidu} but for model $\texttt{nI0.5\_noDC\_l$\zeta$\_noG}$.
\label{conductContributions_lowz}}
\end{figure*}


\begin{figure*}
\includegraphics[width=2.1\columnwidth, clip]{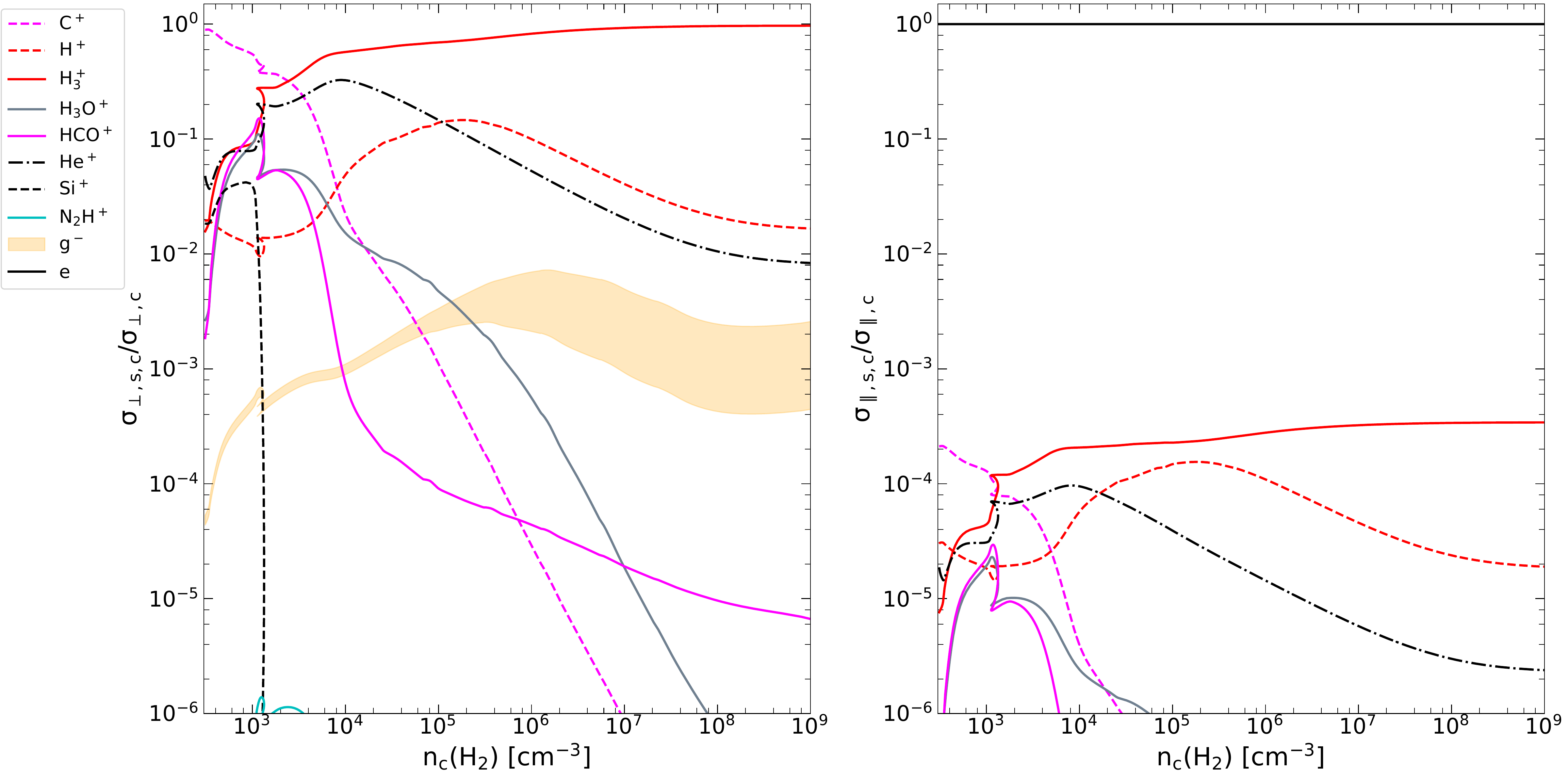}
\caption{Same as in Fig.~\ref{conductContributions_fidu} but for model $\texttt{nI0.5\_noDC\_s$\zeta$\_MRNG}$.
\label{conductContributions_MRN}}
\end{figure*}

From our magnetic models with ambipolar diffusion the model that collapses first is \texttt{nI0.5\_noDC\_l$\zeta$\_noG}, where the cosmic-ray ionization rate is an order of magnitude smaller than the standard value. In fact, in this model, the mass-to-flux ratio exceeds the critical value and the cloud collapses in a timescale that is $\sim$3.2 Myrs shorter than in our fiducial model (\texttt{nI0.5\_noDC\_s$\zeta$\_noG}). This shorter timescale is to be expected, as the number density of charged species in this model is on average a factor of $\sim$ 5 lower than in the fiducial case (see right panel in the second row of Fig.~\ref{molCdensComp} for the density range $10^3$~$\rm{cm^{-3}}$~$\leq$~$n_{\rm{H_2}}$~$\leq$~$10^4$~$\rm{cm^{-3}}$ where the cloud spends most of its evolution) and ambipolar diffusion can proceed faster (see also discussion in Appendix~\ref{vdrift}). This result is also in rough agreement with the ambipolar-diffusion timescale, given by:
\begin{equation}\label{tad}
\rm{\uptau_{AD}} = 1.8\times10^5\Big(\frac{n_i/n_{\rm{H_2}}}{10^{-8}}\Big) yr
\end{equation}
(Ciolek \& Mouschovias 1993) which is valid for the portion of the evolution of the core where $\upmu<\upmu_{crit}$ and from charge neutrality we have that $n_e$ = $n_i$.

From Fig~\ref{CdensTimeEvol}, it can be seen that models \texttt{nI0.5\_noDC\_s$\zeta$\_noG} and \texttt{nI0.5\_DC\_s$\zeta$\_noG}, performed with the small and large chemical networks respectively, do not exhibit any differences for the majority of the time evolution. Only at very late stages, the collapse of the model \texttt{nI0.5\_DC\_s$\zeta$\_noG} is delayed by $\sim$10 kyrs compared to our fiducial model. At first glance, this indicates that deuterium chemistry does not play a significant role in non-ideal MHD effects in molecular clouds. However, as we will show in the next section, deuterated chemical species play, in fact, a dominant role in carrying the conductivities and resistivities, especially at high densities ($n_{\rm{H_2}}\ge10^6~\rm{cm^{-3}}$). This result could also imply that the inclusion of deuterium chemistry is vital when studying non-ideal MHD effects in the protostellar phase.

Finally, Fig~\ref{CdensTimeEvol} reveals little differences between models \texttt{nI0.5\_noDC\_s$\zeta$\_uG} and \texttt{nI0.5\_noDC\_s$\zeta$\_MRNG}, performed considering a uniform and a MRN distribution for the grains respectively, for the majority of their time evolution. The only difference is that model \texttt{nI0.5\_noDC\_s$\zeta$\_uG} reaches a density of $10^9~\rm{cm^{-3}}$ with a delay of $\sim$15 kyrs compared to model \texttt{nI0.5\_noDC\_s$\zeta$\_MRNG}. Compared to our fiducial model, taking grains into account when computing the resistivities increases the ambipolar-diffusion timescale by 1.1 Myrs. Furthermore, as we will show in the next section, the addition of grains leads to significant differences in the Hall conductivity and resistivity, especially at low and intermediate densities.


In the left panel of Fig.~\ref{Brho} we show the evolution of the magnetic field as a function of the central density. In the right panel we show the exponent $\upkappa$ from $B\propto\rho^{\upkappa}$ measured in intervals of half an order of magnitude. Up to densities of $n_{\rm{H_2}}\sim10^4~\rm{cm^{-3}}$, the magnetic field does not scale with density. From densities of $10^4$~$\rm{cm^{-3}}$ the value of the exponent $\upkappa$ is close to the expected value of 0.5 (see Tritsis et al. 2015 and references therein) whereas in the density range $1\times10^6\ge n_{\rm{H_2}}\le 5\times10^7$~$\rm{cm^{-3}}$ we measure $\upkappa = 0.47$, in agreement with Fiedler \& Mouschovias (1993). Interestingly, we find a slightly weaker scaling of the magnetic field with density in model \texttt{nI0.5\_noDC\_l$\zeta$\_noG}. Specifically, considering the density range $1\times10^5\ge n_{\rm{H_2}}\le 1\times10^8~\rm{cm^{-3}}$ we find $\upkappa=0.45$ as opposed to $\upkappa=0.48$ in our fiducial model. This weaker scaling however, can be easily understood from the fact that in model \texttt{nI0.5\_noDC\_l$\zeta$\_noG} there is more diffusion of the magnetic field as a result of the lower cosmic-ray ionization rate.

\subsubsection{Conductivities and Resistivities}\label{condsRes}

In the upper row of Fig.~\ref{conductResi} we show the perpendicular, parallel and Hall conductivities (left, middle and right panels, respectively). With the blue solid and dashed-dotted lines we have overplotted the results by Kunz \& Mouschovias (2010) and Marchand et al. (2016), respectively. We note here that, even though we have matched the initial conditions by Kunz \& Mouschovias (2010) (see section \S~\ref{setup}), we use a standard value for the cosmic-ray ionization rate (exept for model \texttt{nI0.5\_noDC\_l$\zeta$\_noG} where we use $\zeta$/$\zeta_0$ = 0.1). In contrast, Kunz \& Mouschovias (2010) use $\zeta=5\times10^{-17}~\rm{s^{-1}}$ ($\zeta/\zeta_0=3.8$) while Marchand et al. (2016) use $\zeta=1\times10^{-17}~\rm{s^{-1}}$ ($\zeta/\zeta_0=0.8$). Furthermore, we use a constant $\rm{A_v}$ as opposed to both these studies.

For the perpendicular and parallel conductivities, $\sigma_\perp$ and $\sigma_\parallel$, we observe the same trend when we decrease the cosmic-ray ionization rate, as is observed between these two past studies considered here. That is, for number densities $n_{\rm{H_2}}\gtrsim 10^3~\rm{cm^{-3}}$, the two conductivities decrease by a factor of 1.6--2.6 (for $\sigma_\perp$) and $\sim$ 6 (for $\sigma_\parallel$) when a lower cosmic-ray ionization rate is considered (see solid and dash-dotted black lines).

As is evident from the upper right panel of Fig.~\ref{conductResi}, the largest differences between this work and the two previous studies considered, are observed for the Hall conductivity. For the two models where we took grains into account when computing the conductivities and resistivities (\texttt{nI0.5\_noDC\_s$\zeta$\_uG} and \texttt{nI0.5\_noDC\_s$\zeta$\_MRNG}), and for densities $n_{\rm{H_2}}$$<10^5$~$\rm{cm^{-3}}$, the Hall conductivity computed is in line with what was found by Kunz \& Mouschovias (2010) and especially by Marchand et al. (2016). However, as grains start to detach from the magnetic field (see Fig.~\ref{attachFact}), the Hall conductivity is similar to that computed from the models where we have assumed that all grains are neutral. For the simulation where we have assumed a MRN distribution (red dashed line), the deviation in the Hall conductivity commences somewhat later in the evolution since smaller grains detach from the magnetic field at slightly higher densities (see the red shaded region in Fig.~\ref{attachFact}).
 
In the bottom row of Fig.~\ref{conductResi} we show the ambipolar diffusion, Ohmic and Hall resistivities (see Eqs.~\ref{etaadOhm}) as a function of central density. Here, we have overplotted again the results by Kunz \& Mouschovias (2010) and Marchand et al. (2016). The largest differences are observed for the Hall resistivity. At this stage, it is unclear whether these differences are due to the fact that we use a fixed gas-to-dust ratio or because of some other effect. However, one possible explanation could be that we use a much more extended chemical network. Specifically, as we will show below, the most dominant ion is $\rm{H_3^+}$ (see Figs.~\ref{conductContributions_fidu} -- \ref{conductContributions_MRN}). At a density of $10^5$~$\rm{cm^{-3}}$, where the differences in the Hall conductivity begin to occur, the fractional abundance of $\rm{H_3^+}$ in Marchand et al. (2016) is more than 2 orders of magnitude smaller than what is computed here (see Fig.~\ref{molCdensComp} and Fig. 3 from Marchand et al. 2016). Therefore, it is possible that a much heavier ion than $\rm{H_3^+}$ becomes the dominant species and the value of the Hall conductivity does not decrease, even as grains detach from the magnetic field. 

Given that the ratio of the ambipolar-diffusion to Ohmic resistivity is always greater than $10^6$ for the range of densities studied here, ambipolar diffusion is far more important than Ohmic dissipation and the dominant effect responsible for the increase of the mass-to-flux ratio. Only at densities higher than $\sim$ $10^{12}~\rm{cm^{-3}}$ when ions detach from the magnetic field, Ohmic dissipation becomes as, or more important, than ambipolar diffusion (see Tassis \& Mouschovias 2007b).

In each panel in the bottom row of Fig.~\ref{conductResi} we additionally show inset figures in order to better visualize the differences in the resistivities between the models performed with the small and large chemical network ($\texttt{nI0.5\_noDC\_s$\zeta$\_noG}$ and $\texttt{nI0.5\_DC\_s$\zeta$\_noG}$, respectively) for densities $n_{\rm{H_2}}\ge 10^7~\rm{cm^{-3}}$. We find that in the innermost regions and at high densities the ambipolar-diffusion resistivity in model $\texttt{nI0.5\_DC\_s$\zeta$\_noG}$ is 10\% larger than in model $\texttt{nI0.5\_noDC\_s$\zeta$\_noG}$. This not only explains the slightly delayed collapse of the model performed with the large chemical network but is potentially a further indication that Deuterium chemistry might be important for non-ideal MHD effects in the protostellar phase.

In Fig.~\ref{attachFact} we show the ``attachment factor", $\omega_s\uptau_{sn}$ (Tassis \& Mouschovias 2007b), as a function of the central density for the electrons (dash-dotted line), ions (solid lines) and grains (dashed lines and red shaded region). For electrons $\omega_e\uptau_{en}$~$\gg$~1, meaning that they practically remain ``frozen in" to the magnetic field throughout the densities followed here. For heavier ions, the factor $\omega_i\uptau_{in}$ approaches unity for densities $n_{\rm{H_2}}~\ge 10^8$--$10^9$~$\rm{cm^{-3}}$, so it is at this stage that ions begin to detach from the magnetic field. The grains from our simulation where we assumed a uniform distribution detach from the magnetic field for densities $n_{\rm{H_2}}>$$10^5$~$\rm{cm^{-3}}$. Finally, the largest grains from our simulation where we assumed a MRN distribution are never well coupled to the magnetic field since, even at densities of $n_{\rm{H_2}}$ = 300~$\rm{cm^{-3}}$, $\omega_{g^-}\uptau_{g^-n}$~$\ll$~1. On the other hand, smaller grains which are far more abundant, remain coupled to the field up to densities of $10^6$~$\rm{cm^{-3}}$.

In Figs.~\ref{conductContributions_fidu},~\ref{conductContributions_DC},~\ref{conductContributions_lowz} and~\ref{conductContributions_MRN} we present an analysis of the species that carry most of the perpendicular (left) and parallel (right) conductivities for models \texttt{nI0.5\_noDC\_s$\zeta$\_noG}, \texttt{nI0.5\_DC\_s$\zeta$\_noG}, \texttt{nI0.5\_noDC\_l$\zeta$\_noG} and \texttt{nI0.5\_noDC\_s$\zeta$\_MRNG}, respectively. The results from model \texttt{nI0.5\_noDC\_s$\zeta$\_uG} are practically identical to model \texttt{nI0.5\_noDC\_s$\zeta$\_MRNG} and are thus not shown here. In each panel we only show the species that at any given time in the evolution of the core are in the five most important species that carry either the perpendicular or parallel conductivity.

In all models, the species that carries most of the parallel conductivity is the free electrons. Physically, this can be understood from the fact that electrons can stream freely along magnetic field lines. On the other hand, electrons do not contribute significantly to the perpendicular conductivity since they always remain well attached to the magnetic field (see Fig.~\ref{attachFact}) and therefore cannot move perpendicular to the field lines. The contribution of electrons to the perpendicular and parallel conductivities are in excellent quantitative agreement with the results by Kunz \& Mouschovias (2010) (see their Fig. 7). Another common feature across all models is the fact that initially, the species that contributes most to the perpendicular conductivity is $\rm{C^+}$. This is a result of the fact that in our initial conditions, all carbon is in the form of $\rm{C^+}$.

For our fiducial model, we find that the species that carries most of the perpendicular conductivity is $\rm{H_3^+}$, followed by $\rm{H^+}$, in agreement with the results from Tassis et al. (2012b). Although the relative contribution of various species changes, the picture is qualitatively the same for models \texttt{nI0.5\_noDC\_l$\zeta$\_noG} and \texttt{nI0.5\_noDC\_s$\zeta$\_MRNG}. In model \texttt{nI0.5\_noDC\_s$\zeta$\_MRNG}, grain species, specifically in the size range $\sim$0.02--0.03~$\upmu$m also begin to become important at densities $\rm{10^8}$--$\rm{10^9}$~$\rm{cm^{-3}}$, in agreement with the results by Kunz \& Mouschovias (2010). Given that in this model grains become important at late times ($n_{\rm{H_2}}\ge10^6~\rm{cm^{-3}}$), electrons are never part of the five most important species contributing to the perpendicular conductivity at any time during the evolution, and are therefore not shown in the left panel of Fig.~\ref{conductContributions_MRN}. In all these models, $\rm{HCO^+}$ (Desch \& Mouschovias 2001; Tassis \& Mouschovias 2007) only becomes important, if at all, for a very small window in the evolution of the core at densities of $n_{\rm{H_2}}\sim10^3~\rm{cm^{-3}}$. 

The situation drastically changes however, for model \texttt{nI0.5\_DC\_s$\zeta$\_noG} where deuterium chemistry is included. Here, the dominant species, carrying most of the perpendicular conductivity at densities higher that $10^6~\rm{cm^{-3}}$ is $\rm{D_3^+}$, while at even later stages in the evolution of the core $\rm{HD_2^+}$ becomes as important as $\rm{H_3^+}$. Furthermore, in this model $\rm{H^+}$, and at densities $\sim10^5~\rm{cm^{-3}}$ becomes nearly as important as $\rm{H_3^+}$ for the perpendicular conductivity. Similarly to model \texttt{nI0.5\_noDC\_s$\zeta$\_MRNG}, the contribution of electrons to the perpendicular conductivity is not shown in Fig~\ref{conductContributions_DC} since they are never part of the five most important species, at any time during the evolution of the cloud.

\subsubsection{Chemistry Comparison}\label{chemComp}

\begin{figure*}
\centering
\includegraphics[width=2.1\columnwidth, clip]{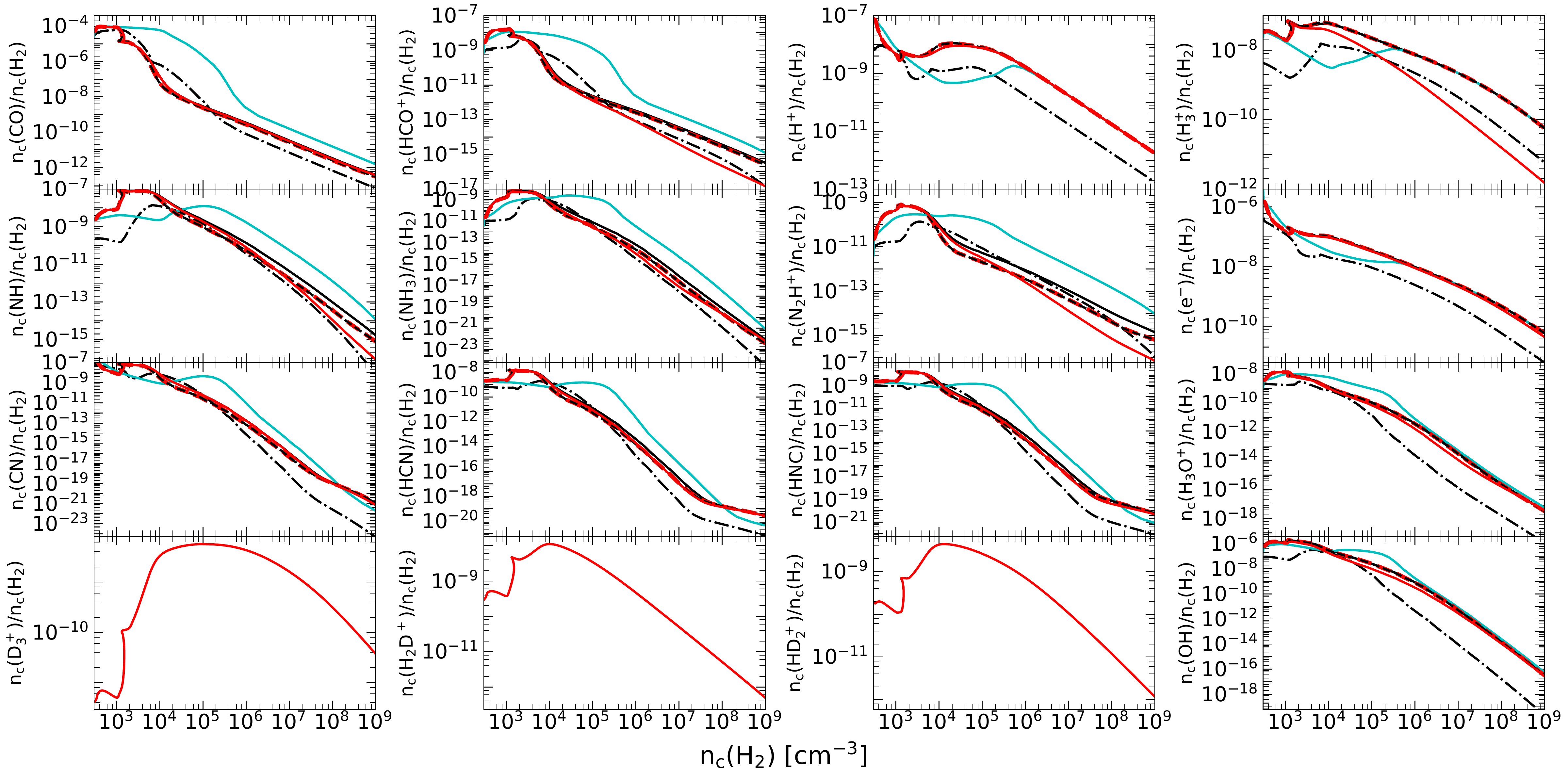}
\caption{Evolution of the abundance of various molecular species as a function of central density from our simulations. Linestyles and colors are as in Fig.~\ref{CdensTimeEvol} and Table~\ref{simsParams}. Interestingly, the abundance of high-density tracing molecules (e.g. $\rm{HCO^{+}}$, CN, HCN, HNC, $\rm{NH_3^{+}}$, and $\rm{N_2H^{+}}$) is over-predicted by one to two orders of magnitude in the hydrodynamical model compared to the magnetic models.
\label{molCdensComp}}
\end{figure*}

The features observed for model \texttt{nI0.5\_DC\_s$\zeta$\_noG} in terms of the conductivities (see also the upper right and lower left panels of Fig.~\ref{molCdensComp}) can be understood considering the main reaction paths for the deuterium enhancement of $\rm{H_3^+}$:
\begin{gather} 
\nonumber \mathrm{H_3^+ + HD \rightleftharpoons H_2D^+ + H_2 + 230 K},
\qquad \\
\nonumber \mathrm{H_2D^+ + HD \rightleftharpoons D_2H^+ + H_2 + 180 K},
\qquad \\
\mathrm{D_2H^+ + HD \rightleftharpoons D_3^+ + H_2 + 230 K}
\label{deuterationPath}
\end{gather} 
(Vastel et al. 2006). For a recent review on the deuterated form of $\rm{H_3^+}$ in prestellar cores from both the theoretical and observational standpoints we refer the interested reader to Caselli et al. (2019). At number densities above $10^3$ -- $10^4$~$\rm{cm^{-3}}$, where CO depletes onto dust grains the destruction of both $\rm{H_3^+}$ and $\rm{H_2D^+}$ in favor of $\rm{HCO^{+}}$ (Pagani et al. 2009) is reduced, leading to a further increase in the abundance of the deuterated forms of $\rm{H_3^+}$.

This interplay can be clearly seen in our chemical network. In Fig.~\ref{molCdensComp} we show the evolution of the abundance of a few basic and commonly-observed species as a function of the central density of the core. Firstly, we observe that the increase in the abundance of $\rm{D_3^+}$ (lower left panel of Fig.~\ref{molCdensComp}) occurs in the same density range as the decrease in the abundance of $\rm{H_3^+}$ (upper right panel). This sharp drop in the abundance of $\rm{H_3^+}$ is only observed for the model \texttt{nI0.5\_DC\_s$\zeta$\_noG} (solid red line) which includes deuterium modeling.

The deuteration of $\rm{H_3^+}$, together with the depletion of $\rm{CO}$ also leads to a decrease in the abundance of $\rm{HCO^+}$ in model \texttt{nI0.5\_DC\_s$\zeta$\_noG} compared to the models that do not include deuterium chemistry (see second panel in the top row). A similar trend to $\rm{HCO^+}$ is also observed for the abundance of $\rm{N_2H^+}$ (third panel in the second row). The production of $\rm{N_2H^+}$ through the reaction of $\rm{H_3^+}$ with $\rm{N_2}$ is reduced leading to a decrease in the abundance of $\rm{N_2H^+}$ in model \texttt{nI0.5\_DC\_s$\zeta$\_noG}. Interestingly, this effect is so pronounced that the abundance of both $\rm{HCO^+}$ and $\rm{N_2H^+}$ at high densities ($n_{\rm{H_2}} \ge 10^6$~$\rm{cm^{-3}}$) is lower in model \texttt{nI0.5\_DC\_s$\zeta$\_noG} rather than in the model \texttt{nI0.5\_noDC\_l$\zeta$\_noG} where the cosmic-ray ionization rate is a factor of 10 lower than the standard value. In Appendix~\ref{spatialChemComp} we additionally present a short comparison of the spatial distribution of the abundances of the species discussed in this paragraph ($\rm{H_3^+}$, $\rm{N2_H^+}$ and $\rm{HCO^+}$) between models \texttt{nI0.5\_noDC\_l$\zeta$\_noG} and \texttt{nI0.5\_DC\_l$\zeta$\_noG}.

Finally, another interesting trend that can be seen from Fig.~\ref{molCdensComp} is that the abundance of several intermediate/high-density tracing molecules, such as $\rm{HCO^+}$, $\rm{NH_3}$, $\rm{N_2H^+}$, $\rm{CN}$ and $\rm{HCN}$, is significantly overpredicted in the hydrodynamical model compared to the models where $\upmu$/$\upmu_{crit}=0.5$. Tassis et al.(2012a) also observed the same behavior regarding the abundance of several commonly observed molecules between their magnetic and non-magnetic models. Similarly, in the simulations presented in Priestley et al.(2019), Priestley et al.(2021) and Yin et al.(2021) the abundances of several molecules were on average a factor of $\sim$2 higher in their models with $\upmu$/$\upmu_{crit}>1$ compared to models with $\upmu$/$\upmu_{crit}<1$ where the duration of collapse is longer.

From the observational standpoint, the picture is less clear. For instance, for L1512 the abundance of $\rm{N_2H^+}$ measured observationally was $\sim$$3\times10^{-11}$ (Lin et al. 2020). This value is more than an order of magnitude lower than what is predicted from our hydrodynamical model at a density of $10^5$~$\rm{cm^{-3}}$. Consequently, a rapid collapse scenario is highly unlikely for this particular core, in agreement with the main results from that study. For the densest parts of B68, Bergin et al. (2002) estimated a $\rm{N_2H^+}$ abundance of $\sim$$3\times10^{-11}$. This value is in better agreement with the results from our magnetic models, although this statement is subject to the estimate adopted for the central density of the core (Alves et al.2001; Roy et al. 2014) and therefore no reliable conclusions can be drawn. Similarly, due to the large observational errors, the abundance of $\rm{N_2H^+}$ obtained for L183 (Pagani et al. 2007) is consistent with the results from both the hydrodynamical and magnetic models. Finally, Chitsazzadeh et al. (2014) observed various molecular spectra towards L1689-SMM16. The abundance they derive for $\rm{NH_3}$ is consistent with the results from our hydrodynamical model. On the other hand, for $\rm{N2H^+}$ and especially $\rm{HCN}$, their results for the same core, are in agreement with the abundances calculated in our magnetic models. This range of possible outcomes when comparing the results of chemodynamical models with observations demonstrates both the need to develop more accurate methods when converting line intensities to molecular abundances as well as the need to better explore the parameter space in simulations.

\section{Summary \& Future prospects}\label{discuss}

We modified the publically-available version of the \textsc{FLASH} AMR code to include a self-consistent calculation of all non-ideal MHD coefficients. Specifically, the perpendicular, parallel and Hall resistivities are computed from the abundances of ionized species as these are computed from non-equilibrium chemical networks. We presented a series of 2D non-ideal MHD simulations of prestellar cores using two of the largest non-equilibrium chemical models to date, and benchmarked our results against previous studies and ideal MHD simulations. 

We found that deuterium chemistry plays a significant role in controlling non-ideal MHD effects in prestellar cores, especially for densities higher than $10^6$~$\rm{cm^{-3}}$. Specifically, we found that the dominant ion and the species that carries most of the perpendicular and parallel conductivities is $\rm{D_3^+}$ followed by $\rm{HD_2^+}$, $\rm{H_3^+}$ and $\rm{H^+}$.

We found that the scaling of the magnetic field with density ($B\propto\rho^{\upkappa}$) is slightly weaker when a very low value is considered for the cosmic-ray ionization rate. Specifically, we find $\upkappa = 0.45$, as opposed to $\upkappa = 0.48$, obtained for the models where we considered a standard value for the cosmic-ray ionization rate. Interestingly, the value of $\upkappa$ obtained for the model where the cosmic-ray ionization rate is just a tenth of the standard value is in better agreement with the value determined observationally for the NGC 6334 and Perseus star-forming regions (Li et al. 2015; Das et al. 2021).

Finally, we found that the abundances of various high-density tracing molecules are overpredicted in the hydrodynamical model, where the core is rapidly collapsing, compared to models where the mass-to-flux ratio was 0.5.

The fact that we are able to follow the evolution of a large number of commonly-observed species opens new pathways for detailed comparisons of such MHD simulations with observations using line radiative-transfer codes (Dullemond et al. 2012; Tritsis et al. 2018). Finally, we presented a new smaller chemical network which consists of 115 species and $\sim$1600 chemical reactions. Despite the fact that this network does not include deuterium chemistry, it leads to a significant increase in computational speed and is thus suitable for performing 3D non-ideal MHD simulations. 

\section*{Acknowledgements}

We thank S. Basu for useful comments and discussions. We also thank the anonymous referee for comments and suggestions that helped improve this manuscript. This work was supported by the Natural Sciences and Engineering Research Council of Canada (NSERC), [funding reference \#CITA 490888-16]. C.~F. acknowledges funding provided by the Australian Research Council (Future Fellowship FT180100495), and the Australia-Germany Joint Research Cooperation Scheme (UA-DAAD). K.~T. acknowledges support by the European Research Council (ERC) under the European Union's Horizon2020 research and innovation programme under grant agreement No. 771282. This research was undertaken with the assistance of resources and services from the National Computational Infrastructure (NCI - grant ek9), supported by the Australian Government. The software used in this work was in part developed by the DOE NNSA-ASC OASCR Flash Center at the University of Chicago. We also acknowledge use of the following software: \textsc{Matplotlib} (Hunter 2007), \textsc{Numpy} (Harris et al. 2020) and the \textsc{yt} analysis toolkit (Turk et al. 2011).
 
\section*{DATA AVAILABILITY}

The data from the simulations presented herewith are available from the corresponding author upon reasonable request.

\appendix
\section{Convergence test}\label{convergence}
In order to test our results for numerical convergence we perform one more simulation using our small chemical network (see Table~\ref{simsParams}) where we decrease the number of blocks by a factor of two such that the initial resolution of our grid is $32\times64$ cells. This simulations is also performed using six levels of AMR refinement and is otherwise identical to model \texttt{nI0.5\_noDC\_s$\zeta$\_noG}.

In Fig.~\ref{converge} we show the central density as a function of time from this simulation (red dash-dot-dotted line). For convenience we have also overplotted our results from model \texttt{nI0.5\_noDC\_s$\zeta$\_noG} (solid black line). The results from these two simulations are identical for the majority of the time evolution of the core. The simulated core where the numerical resolution is a factor of two lower reaches the dynamical collapse phase faster but the difference between these two simulations in terms of the time required to reach this phase is less than 5\%. Consequently, our simulations are sufficiently converged.

As a final test, in Fig.~\ref{IdealMHD100Myrs} we show the evolution of the central density as a function of time from model \texttt{I0.5\_noC\_no$\zeta$\_noG} for 120 Myrs of evolution. Even after this extraordinarily long time, numerical diffusion is not sufficient for the core's mass-to-flux ratio to exceed the critical value and the cloud does not collapse without non-ideal MHD effects included. Instead, the cloud oscillates around its equilibrium state with the oscillations becoming increasingly smaller. Thus, the collapse of the model clouds with $\upmu$/$\upmu_{crit}=0.5$ presented here, is due to physical effects, specifically because of ambipolar diffusion, rather than numerical diffusion.

\setcounter{figure}{0}
\renewcommand{\thefigure}{A\arabic{figure}}

\begin{figure}
\centering
\includegraphics[width=1.\columnwidth, clip]{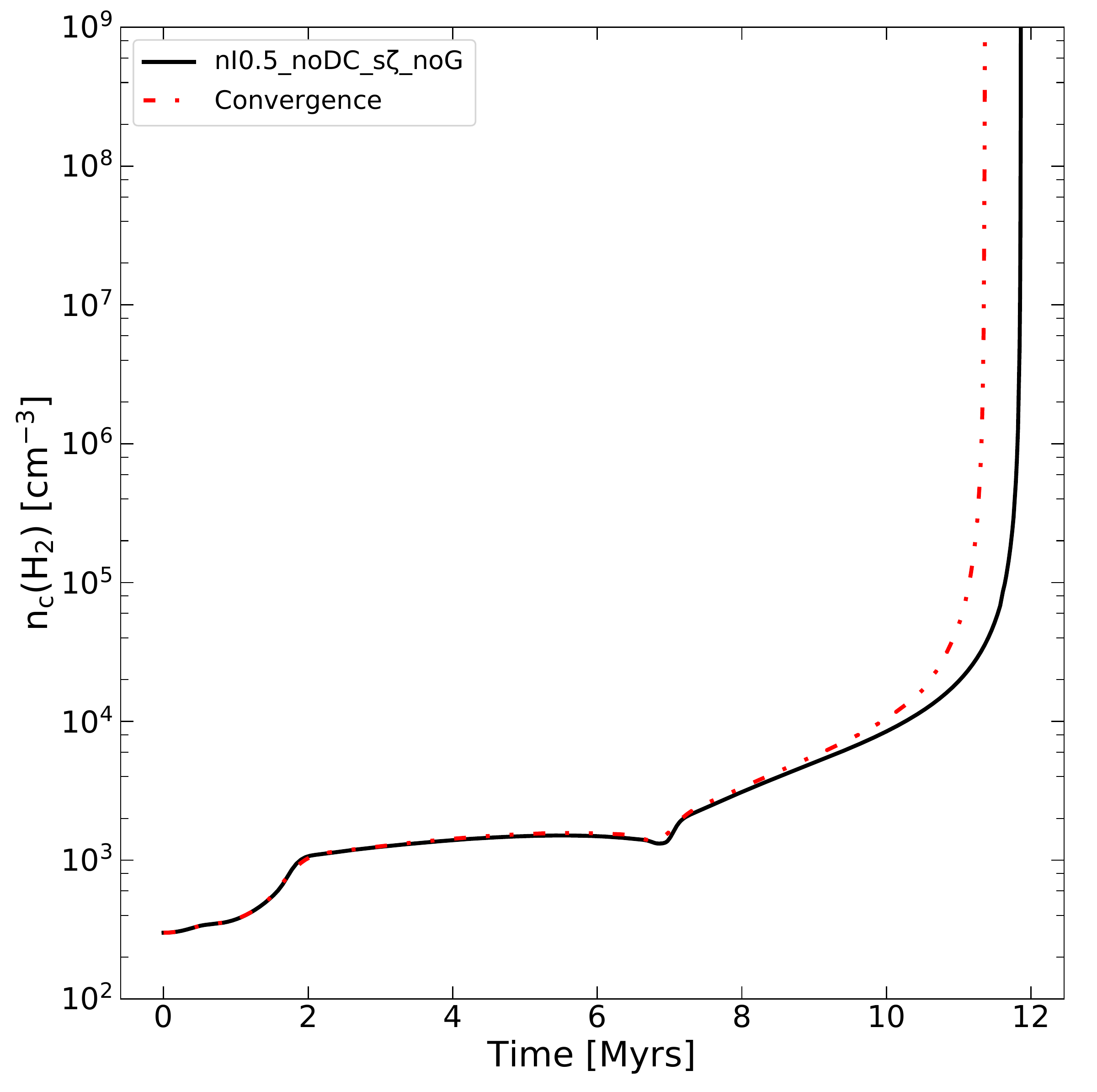}
\caption{Evolution of the central density of the core as a function of time from model \texttt{I0.5\_noC\_no$\zeta$\_noG} (solid black line) and \texttt{I0.5\_noC\_no$\zeta$\_noG\_conv} (red dash-dot-dotted line) where the resolution is a factor of 2 lower.
\label{converge}}
\end{figure}

\begin{figure}
\centering
\includegraphics[width=1.\columnwidth, clip]{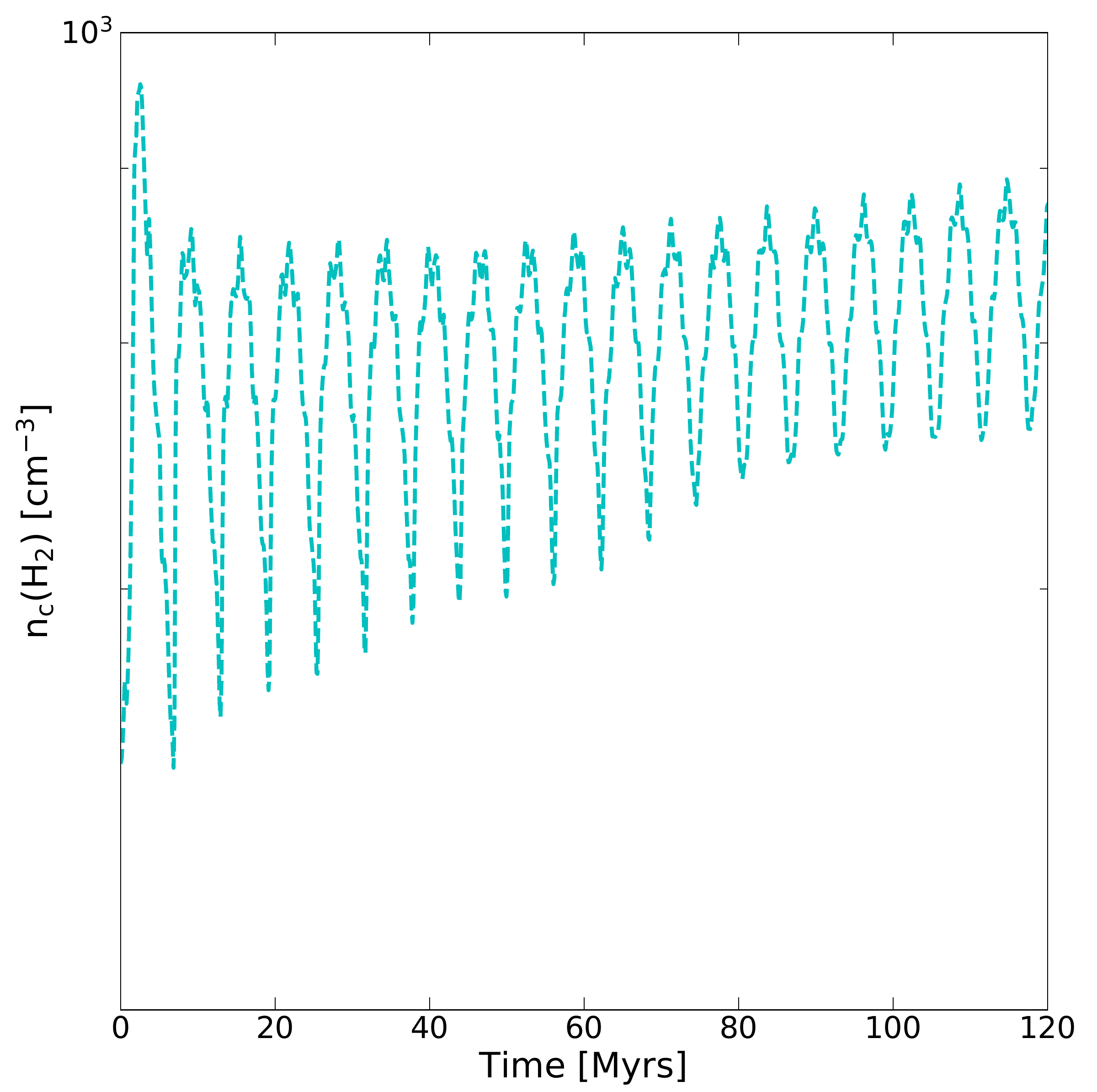}
\caption{Evolution of the central density of the core as a function of time from models \texttt{I0.5\_noC\_no$\zeta$\_noG} where we have left the cloud evolve for 120 Myrs. Without non-ideal MHD effects included, the cloud will not collapse even after a long time period.
\label{IdealMHD100Myrs}}
\end{figure}

\section{Drift Velocities}\label{vdrift}

In this section we calculate the drift velocity between the neutral species and free electrons ($\boldsymbol{v_{drift}}=\boldsymbol{v_n}-\boldsymbol{v_e}$). Given that electrons remain very well attached to the magnetic field lines for the density range studied here (see Fig.~\ref{attachFact}), this drift velocity also corresponds to the drift between the neutral gas and the magnetic field lines. Additionally, since both electrons and ions remain well attached to magnetic field lines for the density range studied here and the Hall term is small compared to the ambipolar-diffusion resistivity, the drift velocity between ions and neutrals is very similar to that between electrons and neutrals.

In Fig.~\ref{driftvelfig} we show median (calculated within $z=\pm$0.1~pc) of the radial drift velocity in units of the sound speed from model \texttt{nI0.5\_noDC\_s$\zeta$\_noG} for each of the snapshots shown in Fig.~\ref{CoreTimeEvolImg}. From Fig.~\ref{driftvelfig} it can be clearly seen that, as the density increases, the drift velocity peaks in progressively smaller radii. Furthermore, for densities $n_{\rm{H_2}}\ge10^4~\rm{cm^{-3}}$ the drift velocity increases from $\sim$0.1 to $\sim$0.3 times the sound speed. For model \texttt{nI0.5\_noDC\_l$\zeta$\_noG} where the degree of ionization is a tenth of the standard value, the drift velocities (not shown here) can be a factor of 2.5--3.5 higher for densities $n_{\rm{H_2}}\ge10^5~\rm{cm^{-3}}$. Such drift velocities could be potentially detectable in spectral-line observations.

\setcounter{figure}{0}
\renewcommand{\thefigure}{B\arabic{figure}}

\begin{figure}
\centering
\includegraphics[width=1.\columnwidth, clip]{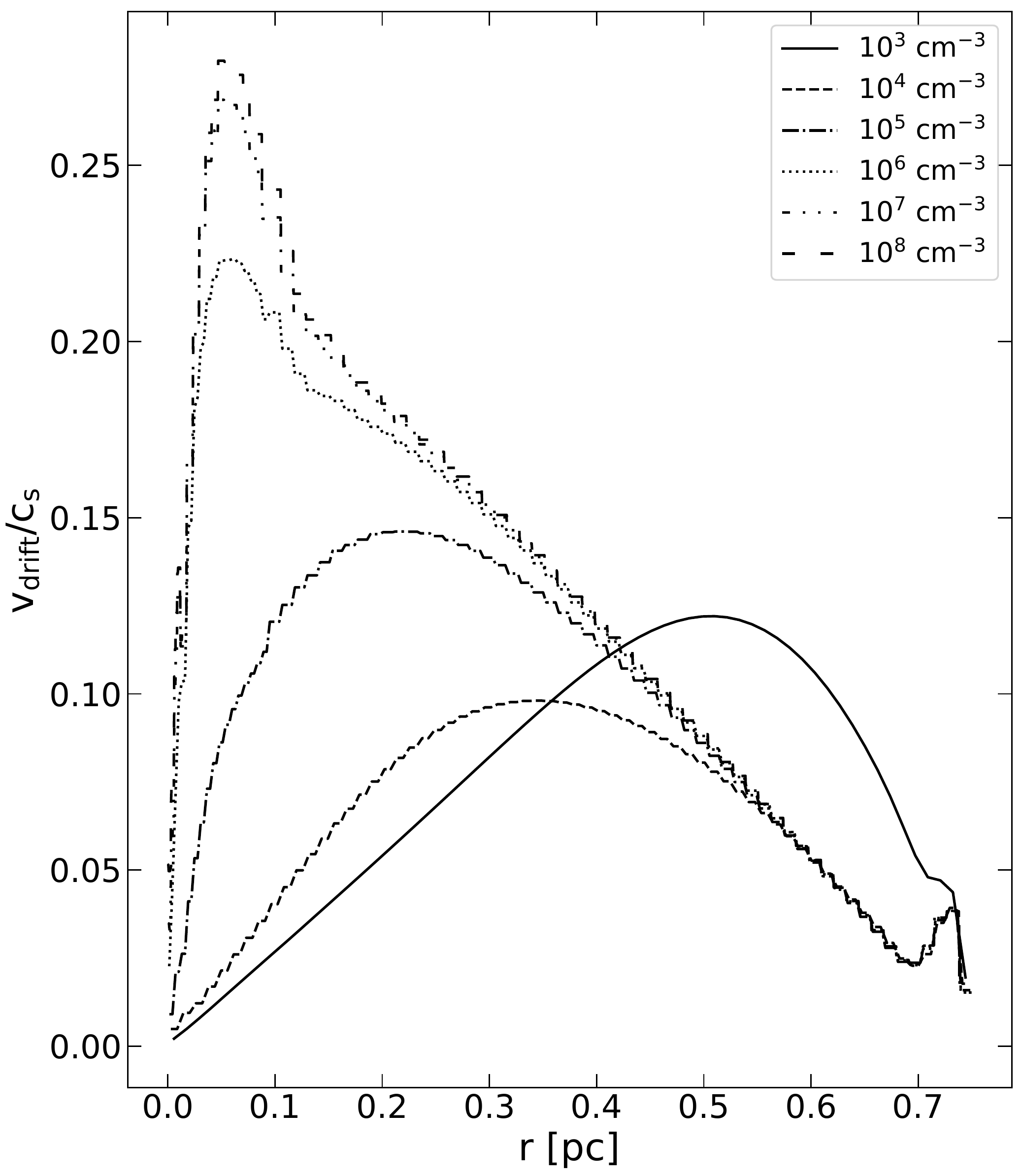}
\caption{Median (within $z=\pm$0.1~pc) of the radial drift velocity in units of the sound speed between electrons and neutral species from model \texttt{nI0.5\_noDC\_s$\zeta$\_noG}. Each line (labeled based on the central number density of the cloud) corresponds to each of the different snapshots shown in Fig.~\ref{CoreTimeEvolImg}.
\label{driftvelfig}}
\end{figure}

\section{Spatial Comparison of Chemical Species}\label{spatialChemComp}

In Fig.~\ref{molCdensComp}, we demonstrated that the abundances of $\rm{H_3^+}$, $\rm{N_2H^+}$ and $\rm{HCO^+}$ exhibit significant differences between models \texttt{nI0.5\_noDC\_s$\zeta$\_noG} and \texttt{nI0.5\_DC\_s$\zeta$\_noG}. In this section, we further compare the spatial distribution of these species between these two models. In the upper row of Fig.~\ref{ChemistryModelComp} we show the abundances of $\rm{H_3^+}$, $\rm{N_2H^+}$ and $\rm{HCO^+}$ (left, middle and right panels respectively) as computed from our fiducial model where we do \textit{not} include Deuterium chemistry. Similarly, in the middle row of Fig.~\ref{ChemistryModelComp} we show the corresponding abundances of these species as computed from model \texttt{nI0.5\_DC\_s$\zeta$\_noG} where Deuterium chemistry is included. In both rows we have zoomed in the innermost 0.1 pc of the cloud and the central number density is $10^9~\rm{cm^{-3}}$.

To further facilitate this comparison, for each of these chemical species, we show in the bottom row of Fig.~\ref{ChemistryModelComp} the ratio of their number density computed from model \texttt{nI0.5\_noDC\_s$\zeta$\_noG} over their number density calculated in model \texttt{nI0.5\_DC\_s$\zeta$\_noG}. From the results presented in Fig.~\ref{ChemistryModelComp}, it is clear that in the densest parts of the core the abundance of these three species is overpredicted by more than an order of magnitude in model \texttt{nI0.5\_noDC\_s$\zeta$\_noG} compared to model \texttt{nI0.5\_DC\_s$\zeta$\_noG}. On the other hand, in the outer parts of the core where the density is $n_{\rm{H_2}}\le10^5~\rm{cm^{-3}}$ (see also the inset figure in the bottom right panel of Fig.~\ref{CoreTimeEvolImg}) the results from these two models are in good agreement.

\setcounter{figure}{0}
\renewcommand{\thefigure}{C\arabic{figure}}

\begin{figure*}
\centering
\includegraphics[width=2.1\columnwidth, clip]{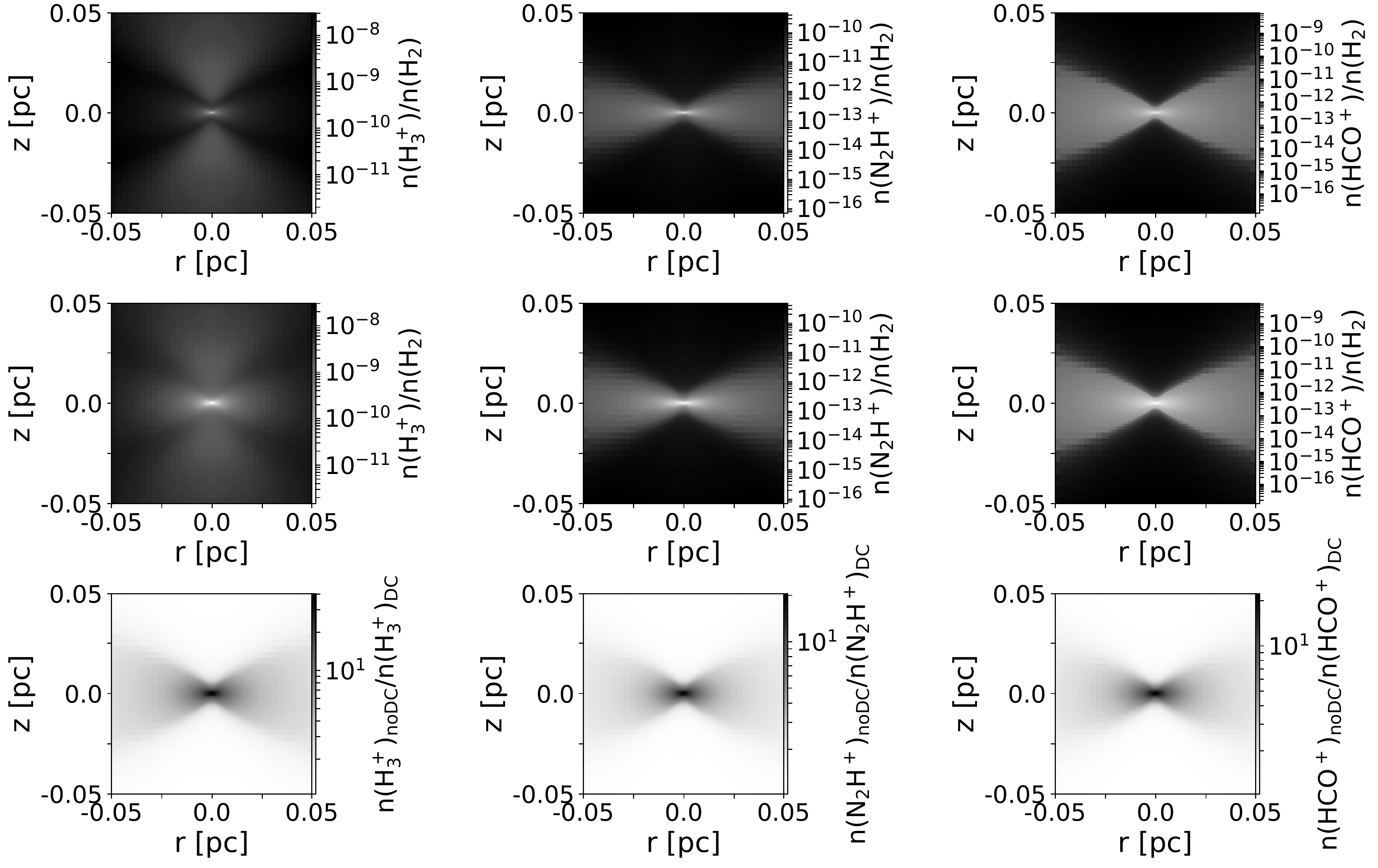}
\caption{Spatial comparison of the abundances of $\rm{H_3^+}$ (left), $\rm{N_2H^+}$ (middle) and $\rm{HCO^+}$ (right) as computed in our model considering our small chemical network (\texttt{nI0.5\_noDC\_s$\zeta$\_noG}; upper row) and the model considering our large chemical network (\texttt{nI0.5\_DC\_s$\zeta$\_noG}; middle row) when the central number density in both models is $10^9~\rm{cm^{-3}}$. To better visualize the differences between the results presented in the upper and middle rows, we additionally show in the bottom row the ratio of number densities of these species between models \texttt{nI0.5\_noDC\_s$\zeta$\_noG} and \texttt{nI0.5\_DC\_s$\zeta$\_noG}.
\label{ChemistryModelComp}}
\end{figure*}

\end{document}